\documentclass[reprint,amsmath,amssymb,aps,showpacs]{revtex4-1}  
\pdfoutput=1

\usepackage{graphicx}
\usepackage{dcolumn}
\usepackage{bm}  
  
\begin{document}

\title{Entropy estimation in bidimensional sequences} 

\author{ F.N.M. de Sousa Filho$^1$,  V. G. Pereira de S\'a$^1$ and E. Brigatti$^2$
}

\affiliation{$1$ Instituto de Computa\c{c}\~ao, Universidade Federal do Rio de Janeiro, Av. Athos da Silveira Ramos, 274, 21941-916, Rio de Janeiro, RJ, Brazil }
\affiliation{$2$ Instituto de F\'{\i}sica, Universidade Federal do Rio de Janeiro, 
Av. Athos da Silveira Ramos, 149,
Cidade Universit\'aria, 21941-972, Rio de Janeiro, RJ, Brazil}
\affiliation{e-mail address: edgardo@if.ufrj.br}

\begin{abstract}

We investigate the performance of entropy estimation methods, 
based either on block entropies or compression approaches, 
in the case of bidimensional sequences.
We introduce a  validation dataset made of 
images produced by a large number of different 
natural systems, in the vast majority characterized by long-range correlations,
which produce a large spectrum of entropies.
Results show that the framework based on lossless
compressors applied to the one-dimensional projection 
of the considered dataset leads to poor estimates. 
This is because higher-dimensional 
correlations are lost in the projection operation.
The adoption of compression methods which do not introduce
dimensionality reduction improves the 
performance of this approach.
By far, the best estimation of the asymptotic  entropy is
generated by the faster convergence of the traditional block-entropies method.
As a by-product of our analysis, we show how a specific
compressor method can be used as a potentially interesting  
technique
for automatic detection of symmetries in textures and images.
\end{abstract}
 

\maketitle


\section{Introduction}


Entropy estimation in bidimensional (2D) systems is a problematic task. 
There is a rich literature studying one-dimensional (1D) systems but 
the analyses of 2D patterns are scarce, and they are expected 
to present new features compared to the 1D case. 
The few works that can be traced \cite{Feldman03,Feldman08,Robinson}, 
written after the 2000's, use techniques based on the block-entropies method, 
inspired by classical theoretical works developed for describing 2D Ising models 
(i.e. in \cite{Alexandrowicz,Meirovitch}).
More recently, a new line of research has brought to the fore 
this problem by rediscovering a well-known 
method based on the use of lossless compression algorithms \cite{Avinery,Martiniani}.
These works have had the merit of showing how this elegant and computationally  
efficient 
technique is a natural candidate for estimating the entropy of 
various 2D systems, including physical ones. In particular, they opened the door 
to estimates based on empirical configurations, and therefore, in essence, images, which 
are a very common result of experimental observations.
Unfortunately, such studies seem to have overlooked
some aspects of the existing literature 
on the use of compressors for entropy estimation. 
In general, the use of these algorithms 
is known to present slow entropy convergence \cite{Plotnik,Plotnik2} 
and alternative, more efficient methods, are 
traditionally used, at least for the 1D case \cite{Grassberger96}.
Furthermore, the existing theoretical results 
which guarantee the convergence of these methods
to the expected entropy value 
refer to specific compressors, called asymptotically optimal algorithms, 
operating on 1D strings \cite{Wyner94}. 
Such results cannot be 
naively generalized to the 2D case. 
In contrast, in the approach applied to the 
2D cases in  \cite{Avinery,Martiniani}, the sequences were trivially projected to 1D 
and then the compressors were applied. 
This projection operation has two important consequences.

First, even if the used compression algorithm for the 1D sequence is optimal, 
it does not imply that 
the  whole approach performs as an asymptotically optimal algorithm,
as the projection to lower dimensionality is part of the compression routine and not just a detail. 
For this reason, the reliability of the method became just heuristic and must be assessed on the basis of empirical tests. 

Second, this  operation, mapping multidimensional patterns to a 
1D sequence loses bidimensional correlations, can be path-dependent 
and even produce spurious long-range correlations \cite{Grassberger86,Feldman03}. 
The use of a locality-preserving curve, like Hilbert's curve, does 
not guarantee to solve these difficulties.
The block-entropies method seems a better candidate to overcome these problems.
In fact, 
is naturally generalizable to patterns of 
higher dimensions  \cite{Politi,Feldman03} 
and the practical implementation of 2D scanning paths
has been successfully tested \cite{Feldman03,Brigatti21}.

Taking into account these considerations,
the aim of this work is to test and compare the accuracy of the two aforementioned methods. 
We will do it by testing the convergence of the two methods to the asymptotic entropy
of different 2D sequences. 
Particular attention must be paid to the choice of the data to be analyzed.
As samples for which entropy estimation is harder are better candidates for the test,
we select systems with long range correlations.
Traditionally, the 2D Ising model with nearest neighbor interactions 
has been considered because of its properties near the critical transition and 
its analytical solution.
Unfortunately, this model is a special case that can be reduced to a 1D string for capturing 
all the statistics that determine its equilibrium properties for entropy estimation, as demonstrated in \cite{Alexandrowicz,Goldstein,Feldman03}. 
For this reason, surely, it is not a good benchmark for entropy estimation in 2D systems. 
Tests of 1D entropy estimation methods have been traditionally carried on using natural systems,
such as written texts and biological sequences. 
Following those approaches, 
in our work we select a large amount of different long-range-correlation natural 
systems presenting a broad spectrum of entropy  values. 

In the following we will discuss in details the properties of this data-set (Section II), the
two considered approaches (Section III) and we present (Section IV) and discuss (Section V) 
the results of our analysis.

\section{Data}

Our pool of data is composed of two datasets
generated by natural systems.
Such systems are characterized by presenting a spontaneous 
superposition of regularities, structures and noises.
Real correlations and noises are much more involved
than synthetic and model-generated ones. In fact,
noise can show intricate variance and correlations 
depending on position and signal. 
The selected data  generally present long-range correlations.
As this property makes the estimation of entropy a particular difficult task 
we can consider this pool of data a suitable benchmark for realizing our test.

Our first sample is composed of 68 built-form maps representing 
urban sections  of cities around the world \cite{Brigatti21}.
A recent analysis has shown that in these systems entropy convergence is similar to the one found in written texts and sonatas, characterized by a slow convergence toward the asymptotic entropy values. 
In fact, the sub-extensive part of their block entropies diverges \cite{Brigatti21,Netto18}, implying the presence of 
subtle, involved correlations at different scales which support entangled long-range structures.

These maps are generated by reducing the urban form to two-dimensional 
arrangements based on building footprints,
which represent the distinctions between built and unbuilt areas. 
Within the considered cities, we focus on small-scale areas with dense 
urban form
and the selection of  sections was based on the identification of 
regions with a high spatial continuity in the fabric of built form. 

\begin{figure}[h]
\centering
\includegraphics[width=0.48\textwidth, angle=0]{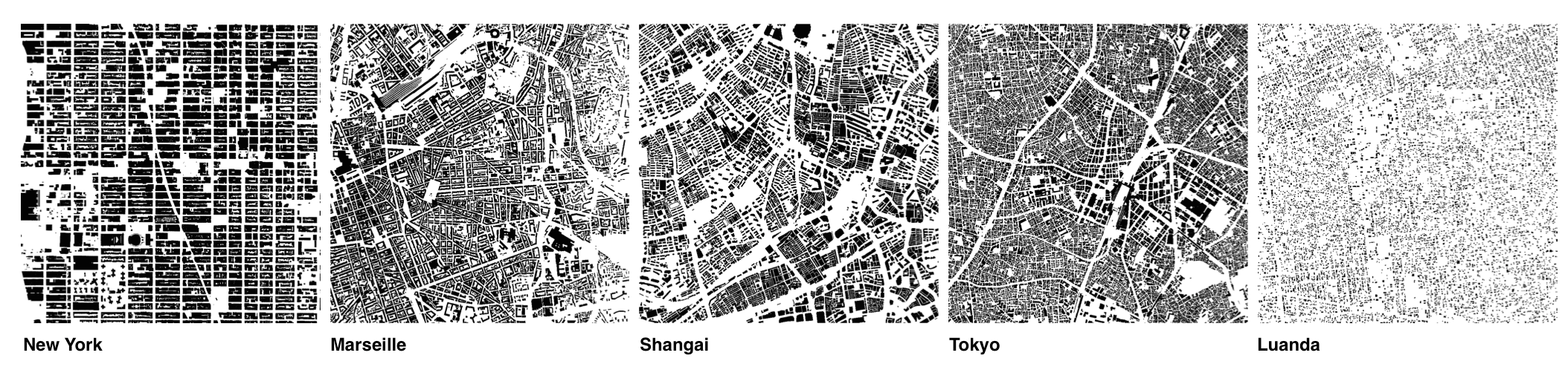}
\includegraphics[width=0.09\textwidth, angle=0]{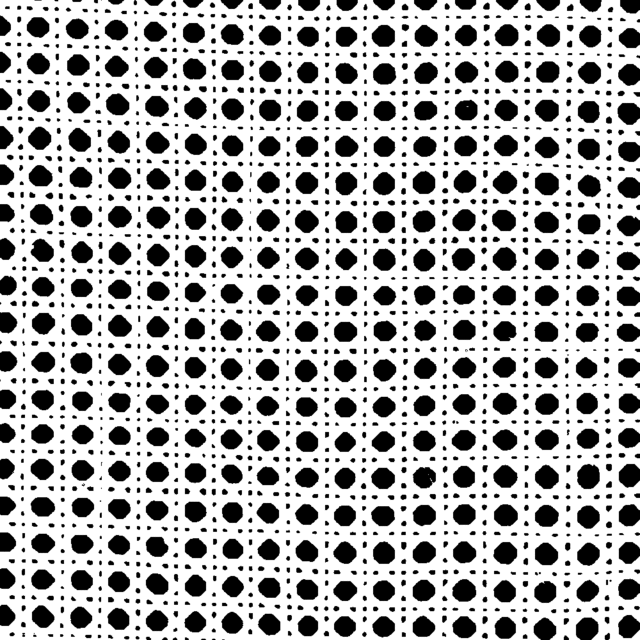}
\includegraphics[width=0.09\textwidth, angle=0]{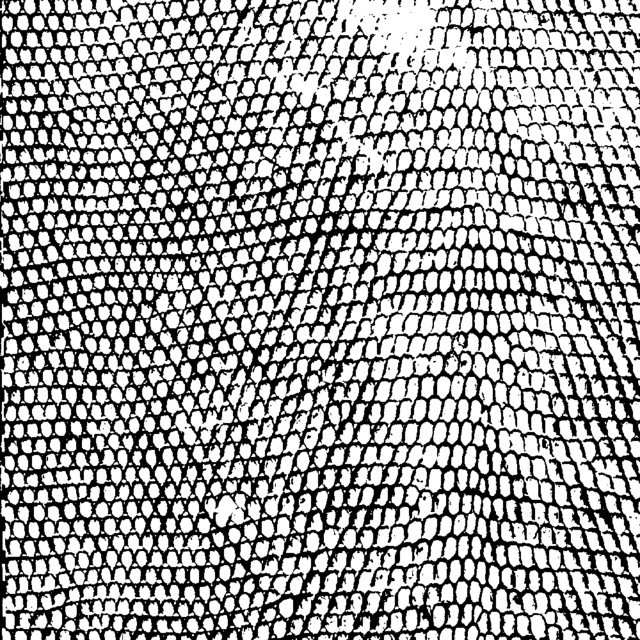}
\includegraphics[width=0.09\textwidth, angle=0]{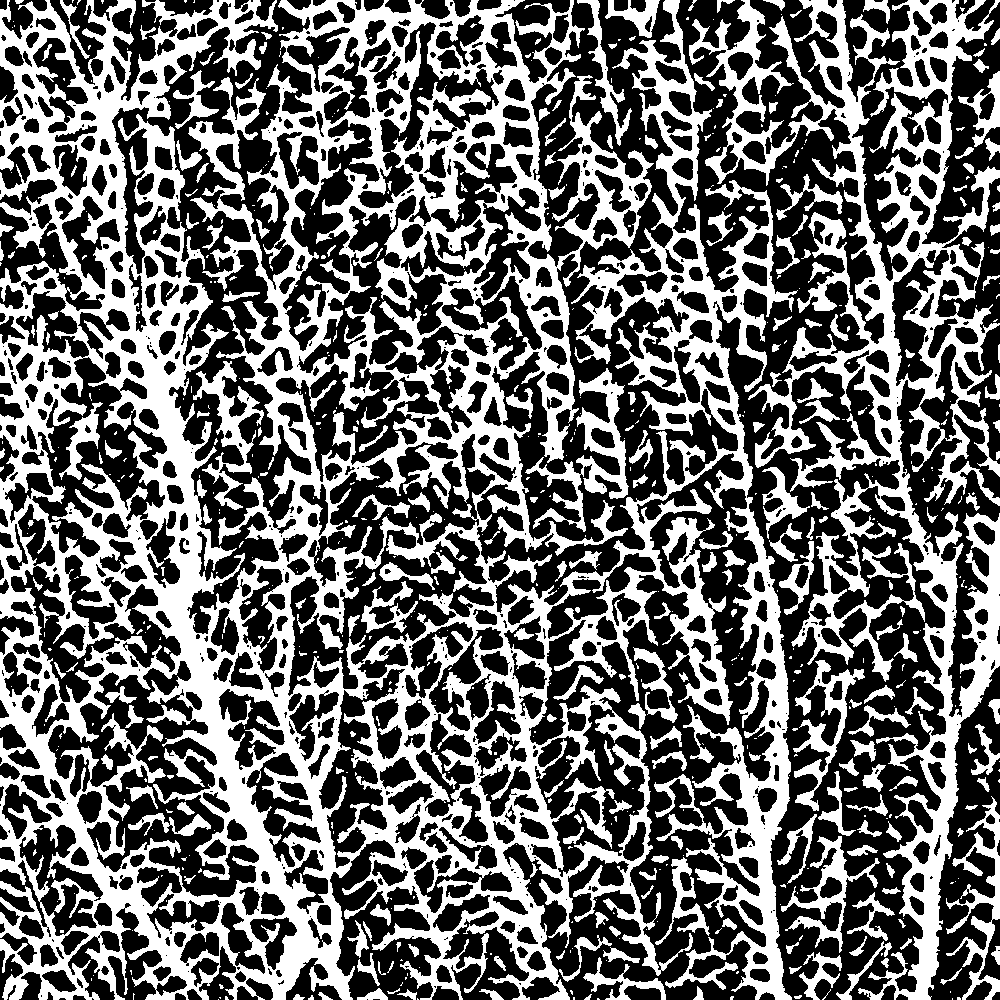}
\includegraphics[width=0.09\textwidth, angle=0]{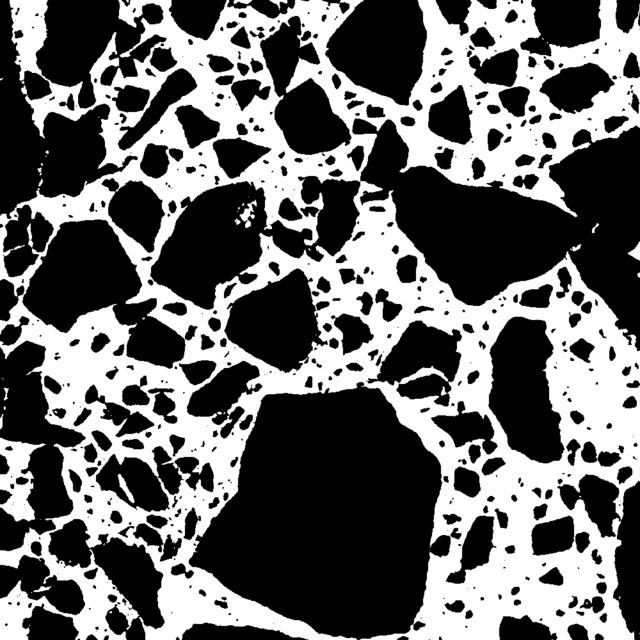}
\includegraphics[width=0.09\textwidth, angle=0]{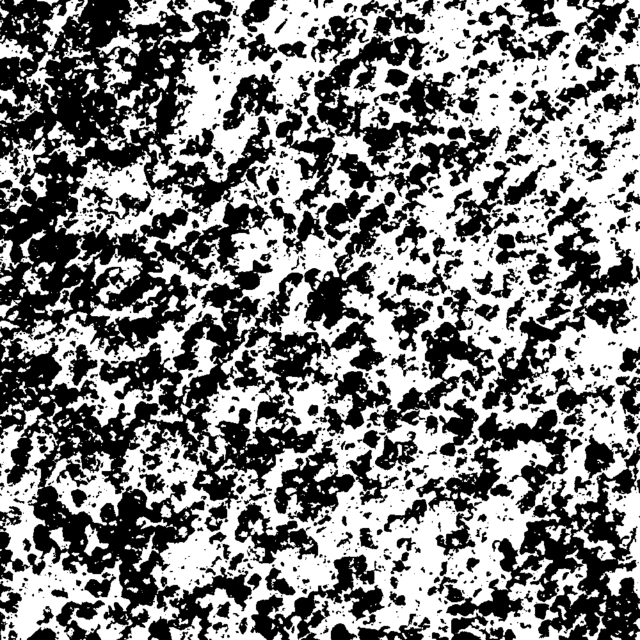}
\caption{\small {\it Top:} Some exemplar maps of urban sections 
in downtown areas.
{\it Bottom:} The binary version of five different Brodatz textures.
} 
\label{fig_data}
\end{figure}

The sample was prepared by extracting 
the data from the public map repository Google Maps 
and selecting geographic areas of $9$ Km$^2$.
Images underwent a re-sizing process, 
were converted to a black-and-white image
and finally into a matrix of size $1000\times 1000$ 
with binary values (1 for built cells, 0 for unbuilt one).
This matrix corresponds to a sequence of $10^6$ cells, 
a size which assures 
a trade-off between typical empirical measurable systems and sufficient statistics
for limiting finite-size effects in the process of entropy estimation.
Figure \ref{fig_data} shows some paradigmatic maps.\\

Our second dataset is composed of Brodatz's textures. 
These images are the most commonly used   
by the signal processing and computer vision communities, 
for validation of techniques of  texture segmentation, classification and image retrieval.
This second dataset presents images generally characterized by long-range correlations,
but  short-range correlations are also present. 
Moreover, we can find more uniform or quasi-periodic textures which are useful
for testing the estimators for entropy values closer to zero. 

The standard Brodatz's texture album \cite{Brodatz}  is composed of 112 grayscale images 
representing 
a reach ensemble of various natural textures, which range from 
simple periodic structures to more random shapes or even relevant noise.
Images were photographed under controlled lighting conditions and present a  very high quality.
As they are not computationally generated, they display natural noises that make 
them better for testing than artificial pictures.
We downloaded  8-bit grayscale images at \cite{download} and we reduced them to binary values,
by using a 50\% threshold to determine whether the resulting pixel would be black or white. 
Finally, we rescaled them to matrices of size $1000\times1000$. 
Figure \ref{fig_Brodatz} shows the binary version of some of the Brodatz's textures.


\section{Methods}

We estimate Shannon's entropy using two methods originally
introduced for 1D systems.

The first method is based on the concept of block entropy.
For 1D sequences, the method consists of defining the block entropy of order $n$ through

\begin{equation}
H_n=-\sum_k  p_n(k) \log_{2}[p_n(k)],
\label{1entropy}
\end{equation}
where blocks are segments of size $n$ of the considered sequence, and the  
sum runs over all the $k$ possible $n$-blocks.
Equation~(\ref{1entropy}) corresponds to Shannon's entropy \cite{Shannon} of the probability distribution $p_n(k)$.
Shannon's entropy 
of the considered system (the whole sequence), which we indicate with $h$, can be obtained by taking the limit for the blocks that goes to infinity \cite{Grassberger96,Lesne}.
This can be done in two ways. The first one is the limit:

\begin{equation}
h=\lim_{n \to \infty} H_n/n,
\label{hentropy}
\end{equation}

which measures the average amount of randomness per symbol 
that persists after all correlations and constraints are taken into account.
The above limit exists for all spatial-translation invariant systems, as demonstrated in \cite{cover}.
Alternatively, the Shannon entropy can be evaluated as the limit of the differential
entropies $h_n=H_n-H_{n-1}$: 
\begin{equation}
h=\lim_{n \to \infty} h_n
\label{hentropy2}
\end{equation}
(note that, for definition $H_0=0$).
This is the limit of a form of conditional entropy, as $h_n$ 
is the entropy of a single symbol 
conditioned on a block of $n-1$ adjacent symbols \cite{Feldman03}. 
The two limits (Eqs. \ref{hentropy} and \ref{hentropy2}) are equivalent. 
More details 
can be found in \cite{Grassberger96,Feldman03,Lesne}.
These approaches can be generalized to sequences of symbols in 2D 
by defining the $n$-blocks for a 2D sequence \cite{Feldman03,Brigatti21}. 

The advantage of using the first limit  (Eq. \ref{hentropy})
is that the set of  $H_n/n$ values is monotonous 
and concave and, in general, displays a clear regularity.
For this reason, when an  
appropriate function for fitting 
the set of $H_n/n$ points can be found, 
the limit 
can be empirically obtained  estimating its asymptote.
Unfortunately, the form of the convergence of the $H_n/n$ 
is not universal. It depends on the 
behavior of the correlations present in the analyzed system 
and it is not possible to define an universal unsatz. 

In contrast, generally, the second approach of equation \ref{hentropy2} is more 
influenced by statistical errors, which do not suggest the use of 
a fitting function for extrapolating the limiting value of $h$,
but it presents a really faster convergence \cite{Grassberger96}.
As we are going to estimate entropy for a varied pool of data, 
which displays very different systems with distinct correlation structures, 
we will use this second approach, which take the limit of the differences.
The definition of the $n$-blocks for bidimensional sequences 
will use the 2D blocks defined by Feldman {\it et al.} in \cite{Feldman03}, which
are inspired by classical analysis of Spin systems. 
This block-entropies estimation method has already been applied in previous works,
where the robustness and reliability of the method has been shown also
for bidimensional systems  \cite{Feldman03,Feldman08,Robinson,Netto18,Brigatti21}.\\


The aforementioned Shannon's approaches for entropy estimation 
are based on probabilistic concepts referring to the source that emits the set of 
all possible sequences.
In contrast, it is possible to estimate the entropy on the basis of ideas
defined for a single finite sequence.

The basic concept is the algorithmic (or Kolmogorov) complexity (AC) 
which measures the complexity of an individual object
by the size of the smallest program that can reproduce it. 
In fact, the AC  of a sequence $x$ ($C(x)$) is the length of the shortest 
program which generates as output the sequence and stops afterwards \cite{Kolmogorov,Chaitin}.
 For any  probability distribution $P(x)$ that is computable using a Turing machine (a very general condition), 
 the expected value of AC equals Shannon's entropy, up to a constant term \cite{Li08}.
From this result it follows that Shannon's entropy  is asymptotically equal 
to the expected complexity: $\sum_x P(x)C(x)\sim -\sum_x P(x)logP(x)$ \cite{Li08}.


Unfortunately, the application of this powerful concept for the estimation of $h$
is limited by the fact that AC is a non-computable quantity.
Even if the minimal theoretical program that generates the sequence is not
achievable, there are compression algorithms which can over-approximate it. 
Among them, asymptotically optimal algorithms are the ones for which
the ratio of the length of the compressed  and uncompressed files
tends to $h$ when the length of the sequence tends to infinity.
For sequences  emitted by finite-alphabet stationary ergodic sources,
a famous optimal algorithm for 1D 
sequences is the Lempel-Ziv algorithm (LZ77) \cite{LZ77}.
The convergence to $h$ is slow, with corrections behaving 
like $O(\frac{loglog(N)}{log(N)})$ \cite{Wyner94}. 

An interesting implementation of these ideas was recently introduced
by Avinery {\it et al.} in \cite{Avinery}, in a framework 
that can be naturally extended also to
sequences of continuous values and  bidimensional data.
In such heuristic approach, entropy is evaluated following this scheme: 

1) discretize the considered configurations 

2) store them in a 1D file 

3) measure the compressed file size with a  lossless compression algorithm ($C_d$)

4) estimate the incompressibility $\eta$ by:
$\eta=(C_d-C_0)/(C_1-C_0)$
where $C_0$ is a compressed  degenerate  datasets   and $C_1$ a compressed random
dataset

5) map $\eta$ to the asymptotic entropy $h$: $h=\eta h_{max}$

Here we use this scheme for binary variables.
Note that, even if the used compression algorithm for the 1D sequence is optimal,
it  does not guarantee that the framework works as an asymptotically optimal algorithm.
In fact, before applying the compressor,  the algorithm 
reduces the bidimensional sequence to a 1D one.
For this reason, in this work we will test the implementation of this framework
for general lossless compression algorithms, not just LZ77, considering also methods 
which operates directly on 2D systems, without previously reducing the system to a 1D sequence. 
To sum up, for our study, the algorithm reduces to measure the compressed file size with 
a  lossless compression algorithm    
and estimate the incompressibility $\eta$, which corresponds to the asymptotic entropy $h$.


We will consider six lossless compression algorithms, all implemented using Python Libraries.
A first group of algorithms are based on the classical LZ77 algorithm \cite{LZ77},
which compresses one-dimensional strings, detecting repeated substrings 
and replacing them with pointers to a dictionary. 
Among state-of-the-art compressors using the LZ methods, one is Gzip \cite{Gzip}.
As it is not constructed for image compression, the images were firstly linearized using Hilbert's curves, which preserve locality and are suggested to produce the best performance in 2D \cite{moon}, and then compressed.
The second algorithm, Png \cite{Png}, uses a compression method similar to Gzip, but already adapted for images. For each line, it applies a filter that turns colors into color differences, and then compresses that line using a combination of LZ77 and Huffman coding. 
Gif \cite{Gif_all} is a popular format that allows animations, and its compression is lossless as long as the original image has only 8 bits of color.   Images are compressed using the Lempel-Ziv-Welch algorithm \cite{LZW}, an improved implementation of the LZ78 algorithm \cite{LZ78}.
WebP \cite{Gif_all} is by default a lossy format, but it can be used in a lossless mode. 
It works by applying a series of reversible filters to the image that make it more compressible, and then replacing each repeated horizontal pixel sequence with a reference to where it previously appeared, implementing a variation of the LZ77 method.

The last two compressors, Jpeg-2000 \cite{Gif_all} and Jpeg-ls \cite{Gif_all}, 
are not based on LZ algorithms and
are variations of Jpeg, a popular lossy compressor. 
Jpeg-2000 is used in lossless compression mode. 
It splits the image in so-called tiles which are wavelet transformed and finally encoded.
The algorithm used by Jpeg-ls is LOCO-I \cite{jpeg-ls}, a dynamic compression algorithm 
that uses statistical inferences to realize a lossless compression.
The algorithm consists of three parts, performed for each pixel. First the prediction is done, then the context is determined, and finally the coding is completed. The first step, the prediction, is done by trying to estimate what would be the target pixel value based on 3 neighbors positioned around it.
From the estimate, the forecast error is calculated and then corrected, making use of the context values. 
Finally, the error is compressed using the Golomb-Rice coding.
The scan format of this method has some analogies with the format used in the entropy estimation via differential entropy. 

In the case of Gzip, 
the matrix format is passed to the compressor after linearization.
In the cases of Png, Gif, WebP, Jpeg-2000, and Jpeg-ls the matrix is transformed 
into a bitmap (it just imply the color black pixel to be numbered 0 and the white 255, instead of 1) 
and subsequently compressed.

More details describing the block-entropies and the compression methods used in our analysis can be found 
in the Appendix.

\section{Results}

\begin{figure}[h]
\begin{center}
\vspace{0.6cm}
\includegraphics[width=0.2\textwidth, angle=0]{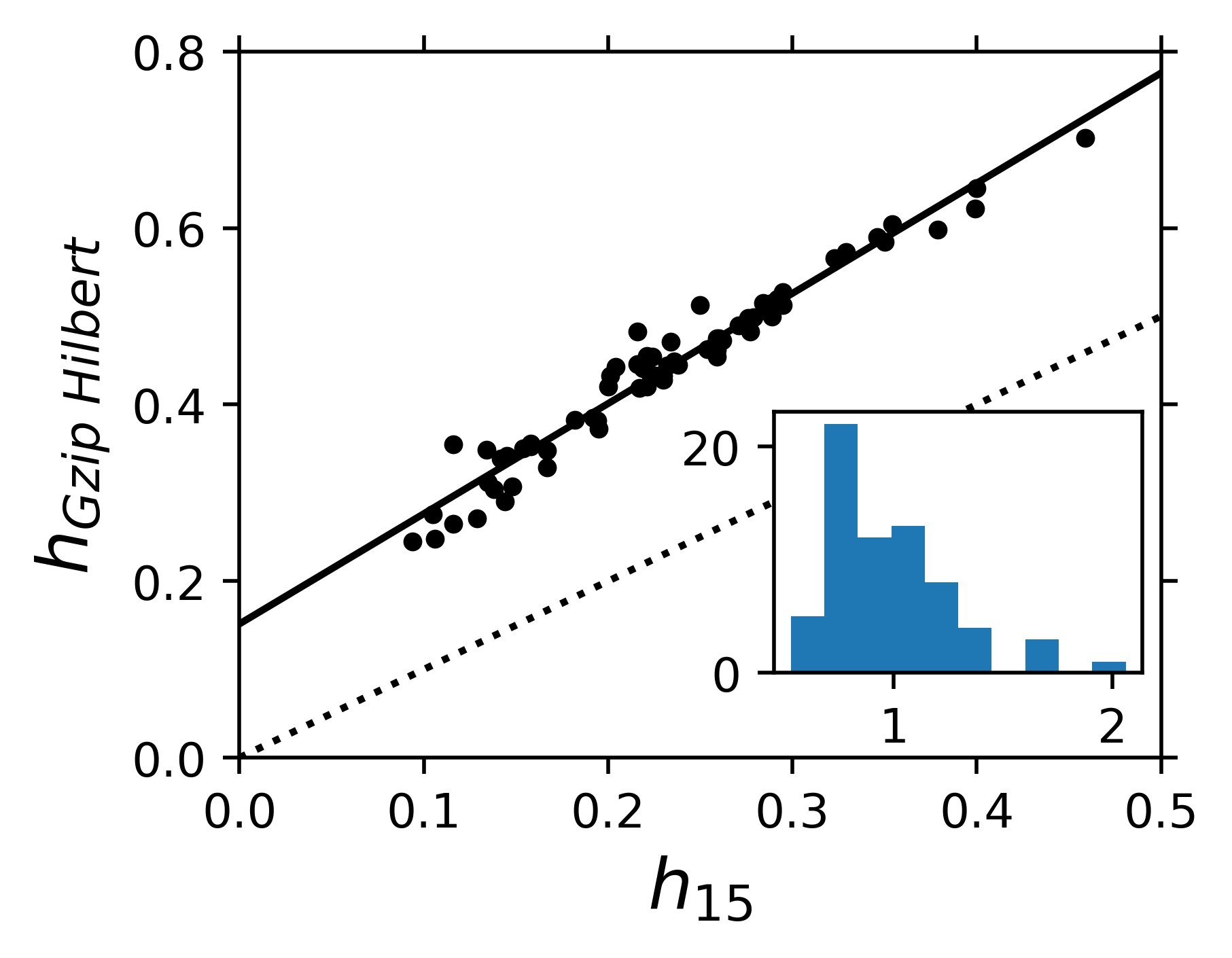}
\includegraphics[width=0.2\textwidth, angle=0]{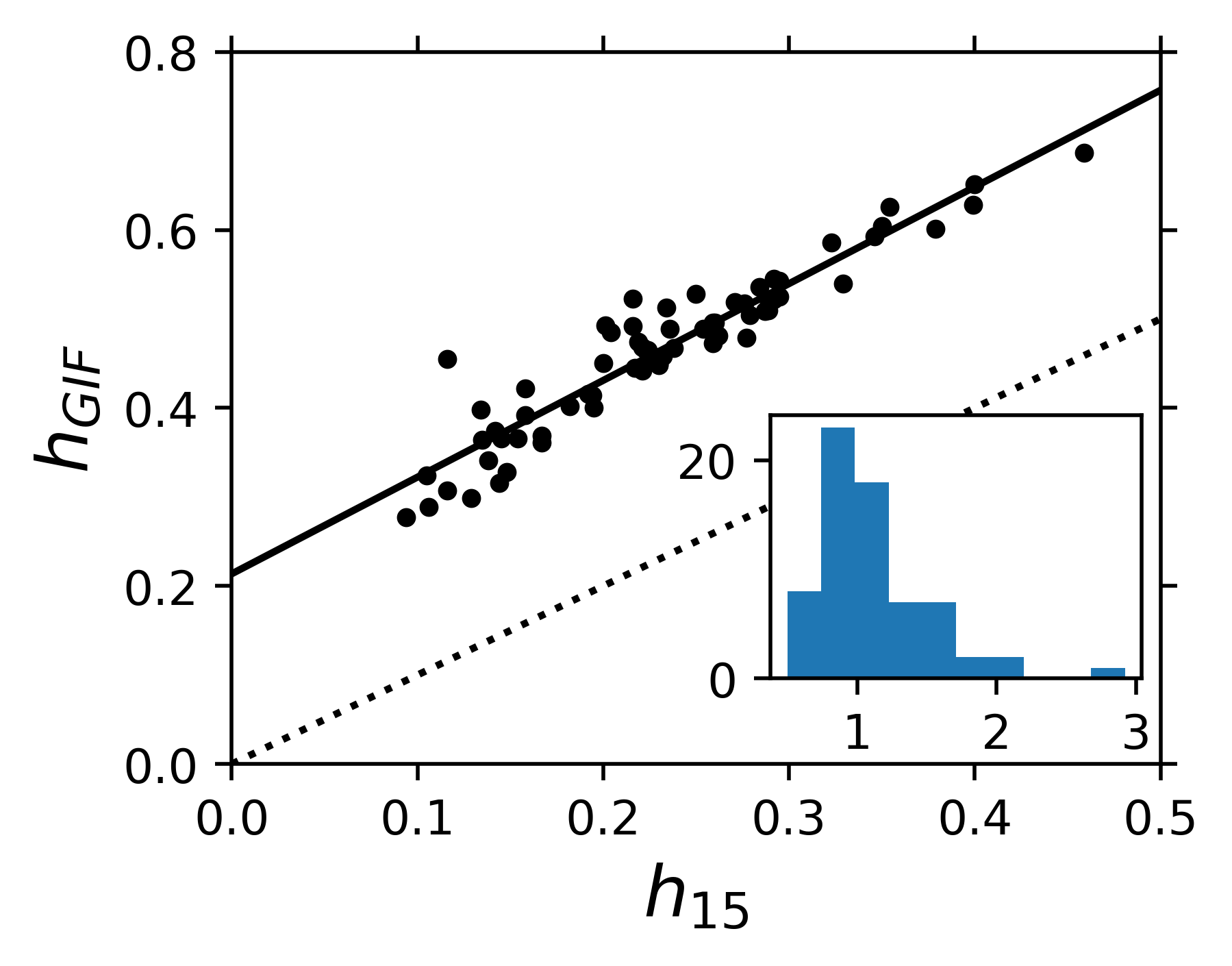}
\includegraphics[width=0.2\textwidth, angle=0]{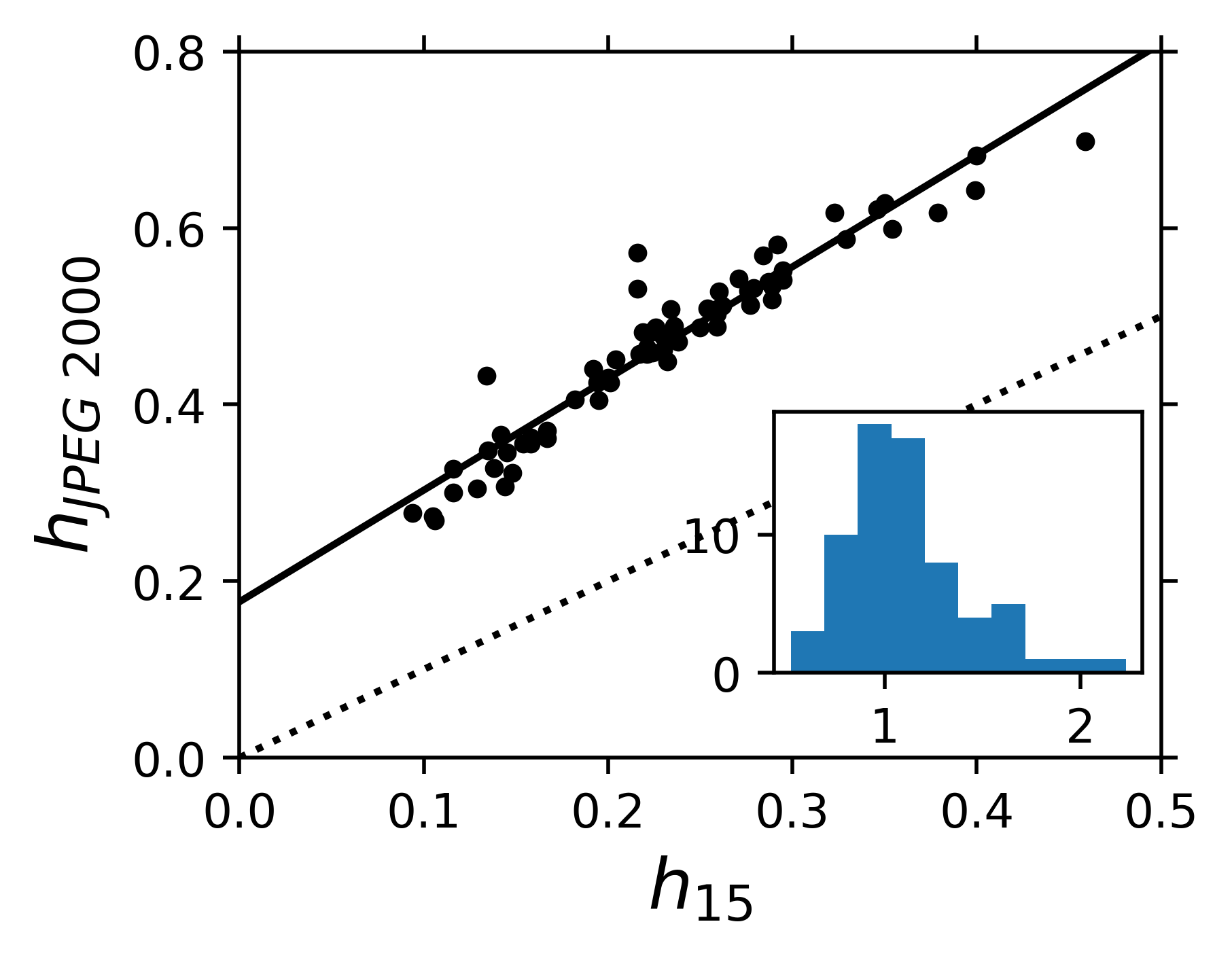}
\includegraphics[width=0.2\textwidth, angle=0]{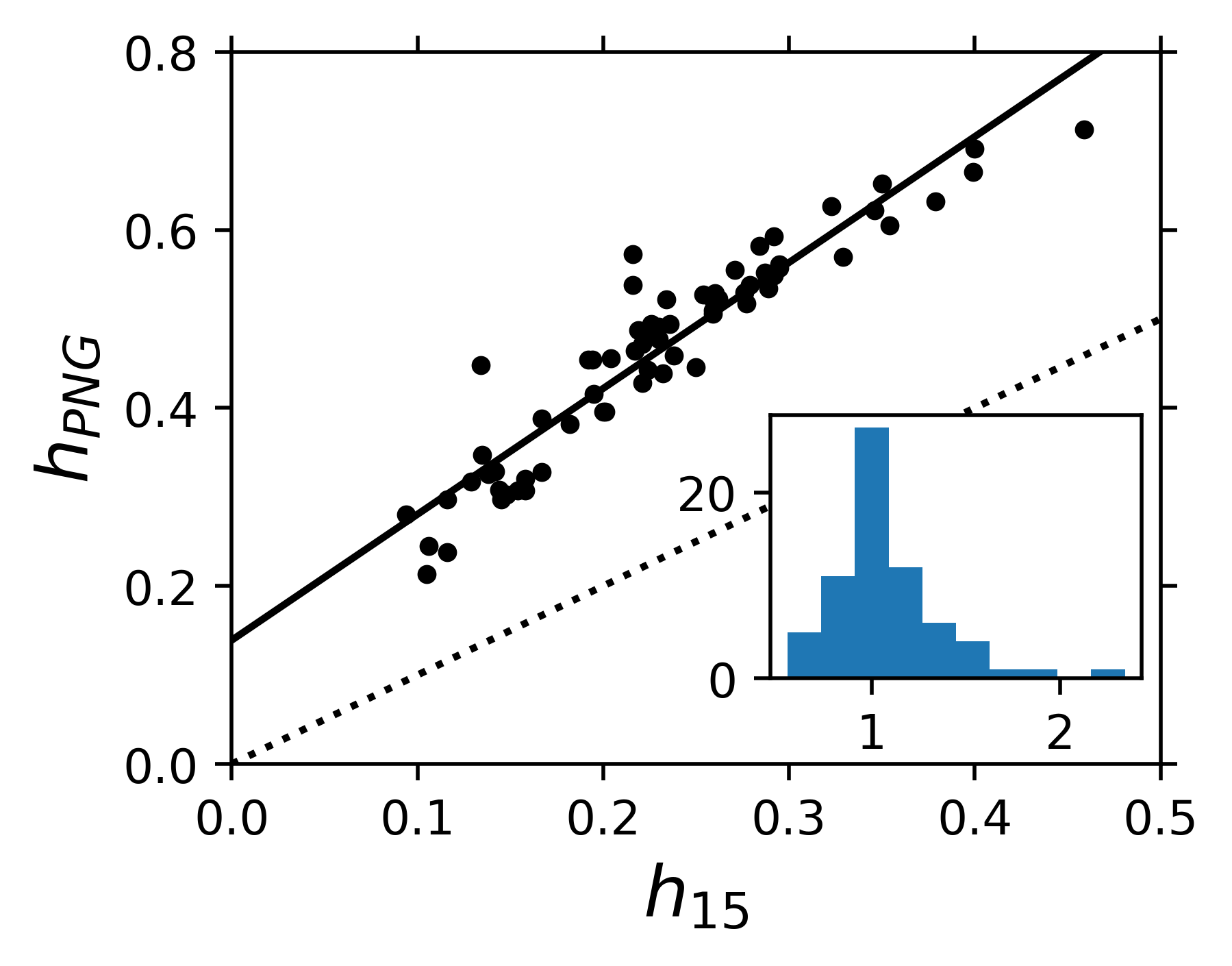}
\includegraphics[width=0.2\textwidth, angle=0]{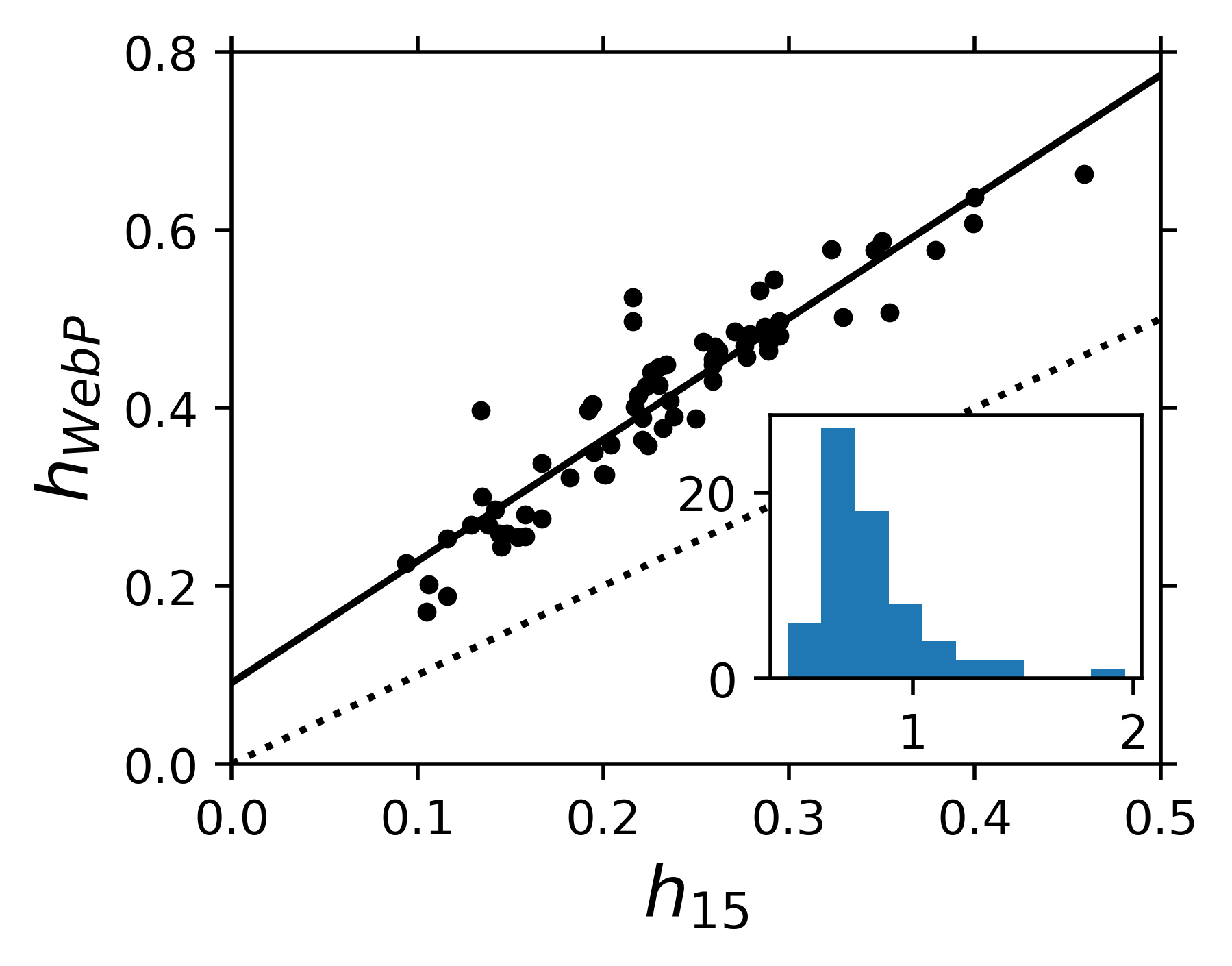}
\includegraphics[width=0.2\textwidth, angle=0]{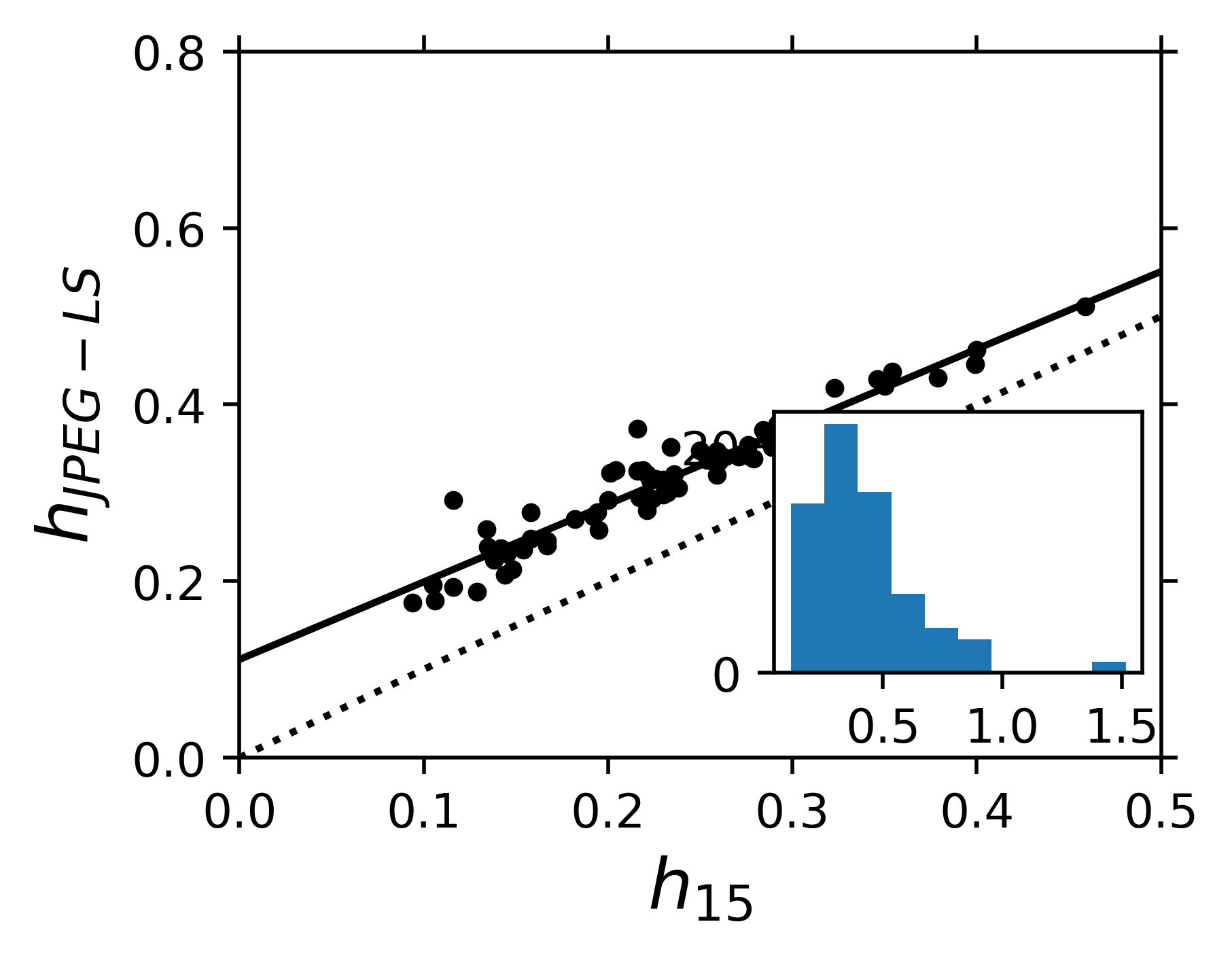}
\end{center}
\caption{\small Scattering plot of the entropies of the urban sections evaluated 
using the lossless compression algorithms versus the block-entropies algorithm. 
Each figure represents the entropy pairs obtained using a different compressor. 
Continuous lines represent  the linear fitting of the points. In the inset,   
the histogram of the relative differences (equation \ref{eq_diff}) 
between the two considered methods.}
\label{fig_city}
\end{figure}

\begin{figure}[h]
\begin{center}
\vspace{0.6cm}
\includegraphics[width=0.2\textwidth, angle=0]{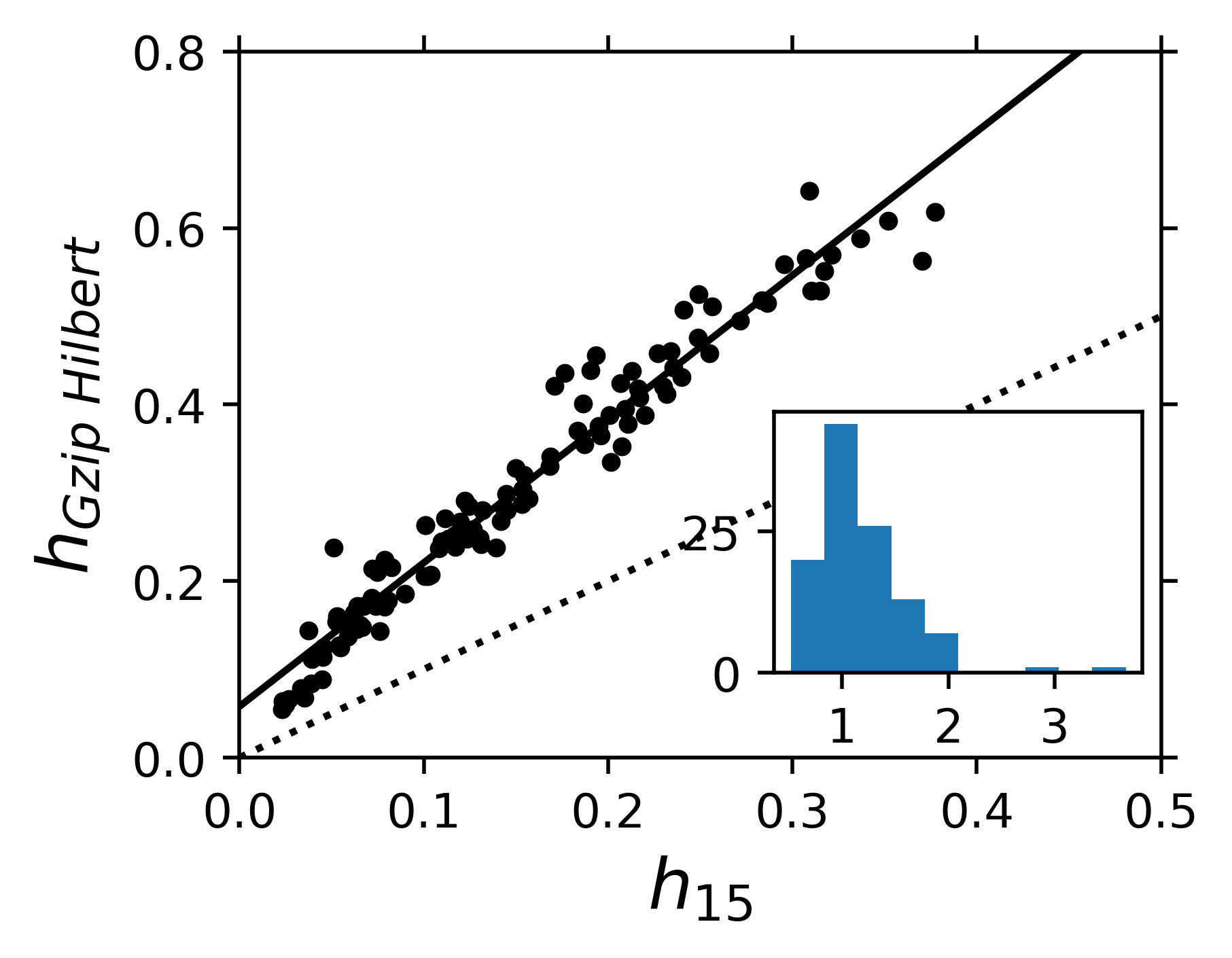}
\includegraphics[width=0.2\textwidth, angle=0]{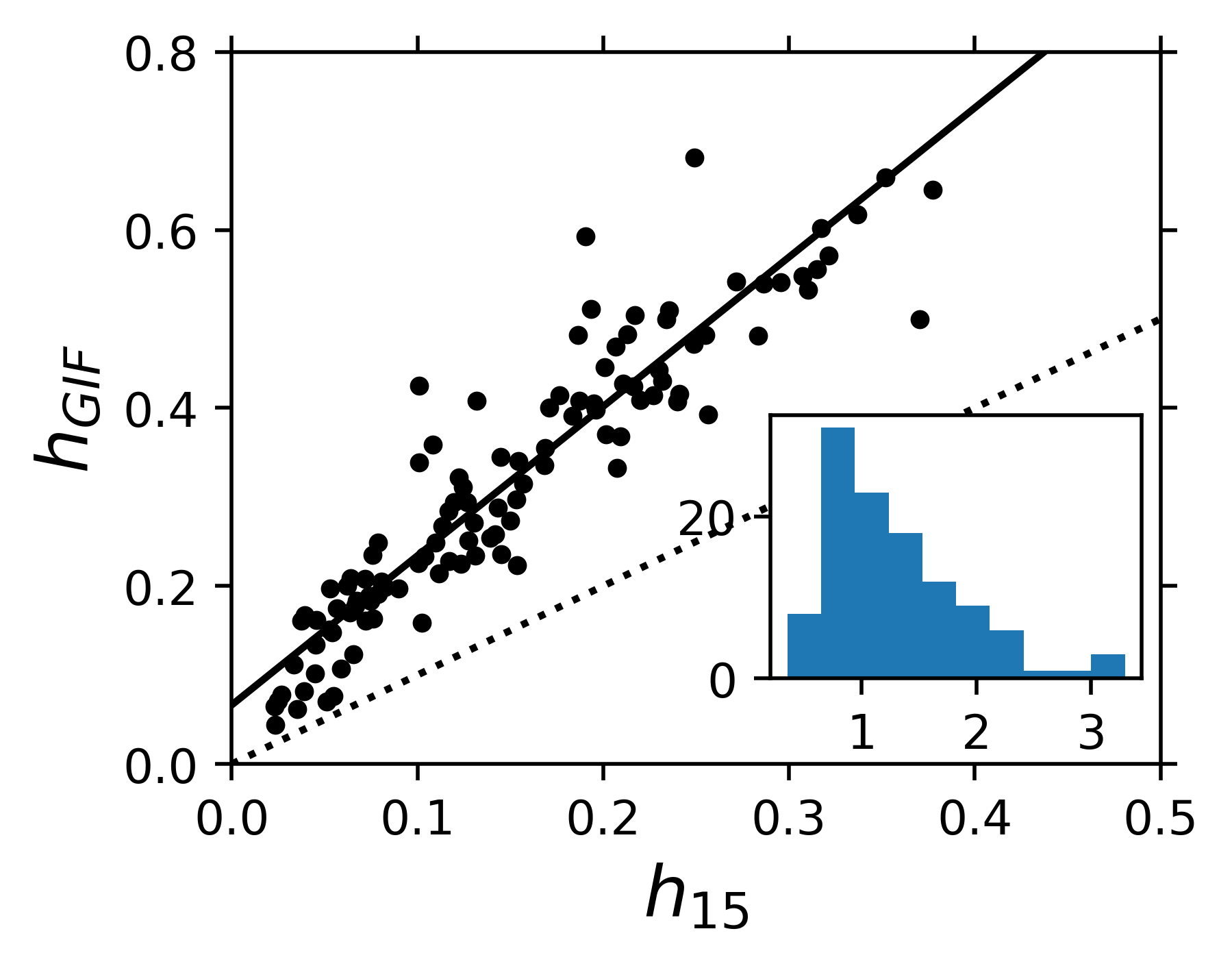}
\includegraphics[width=0.2\textwidth, angle=0]{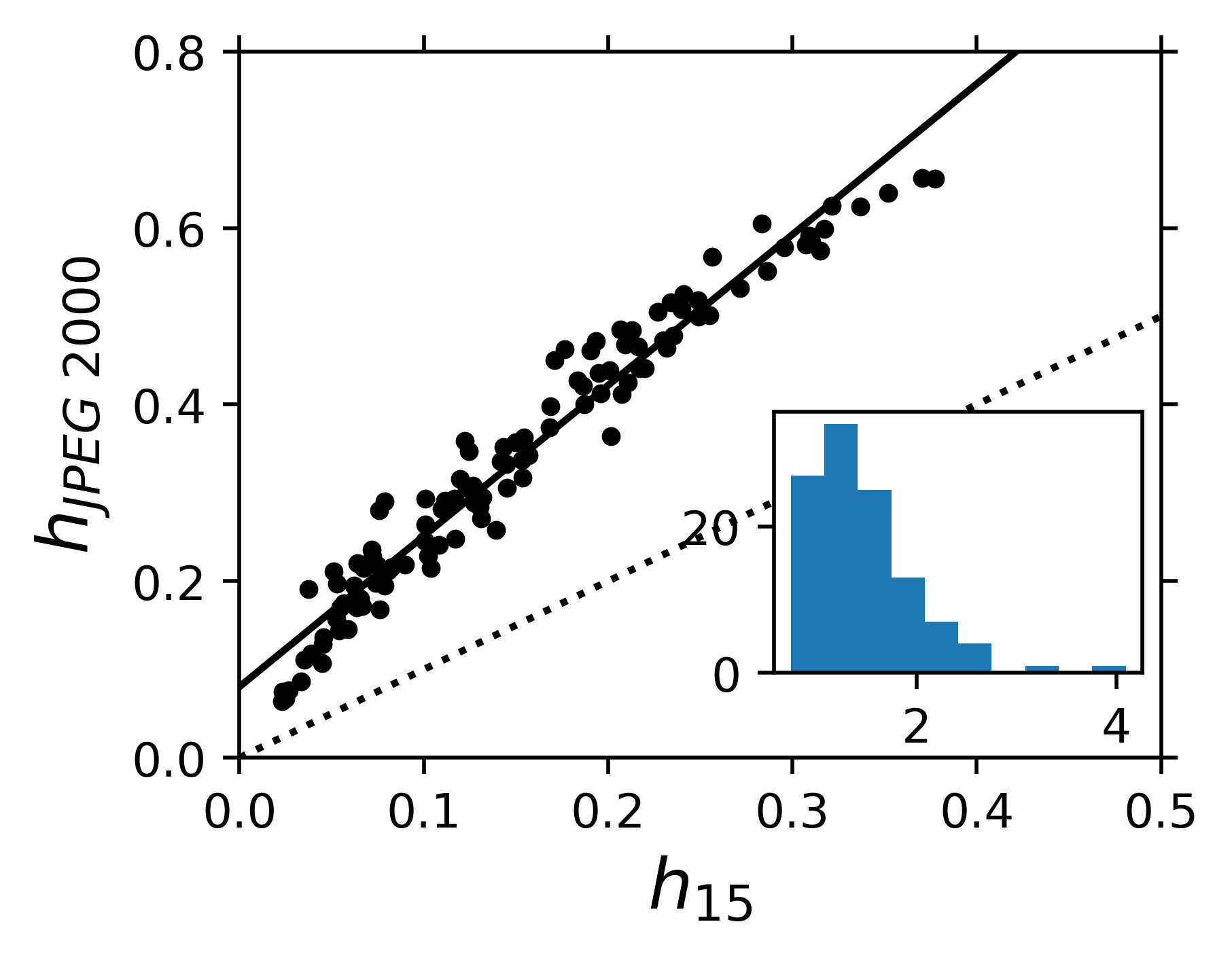}
\includegraphics[width=0.2\textwidth, angle=0]{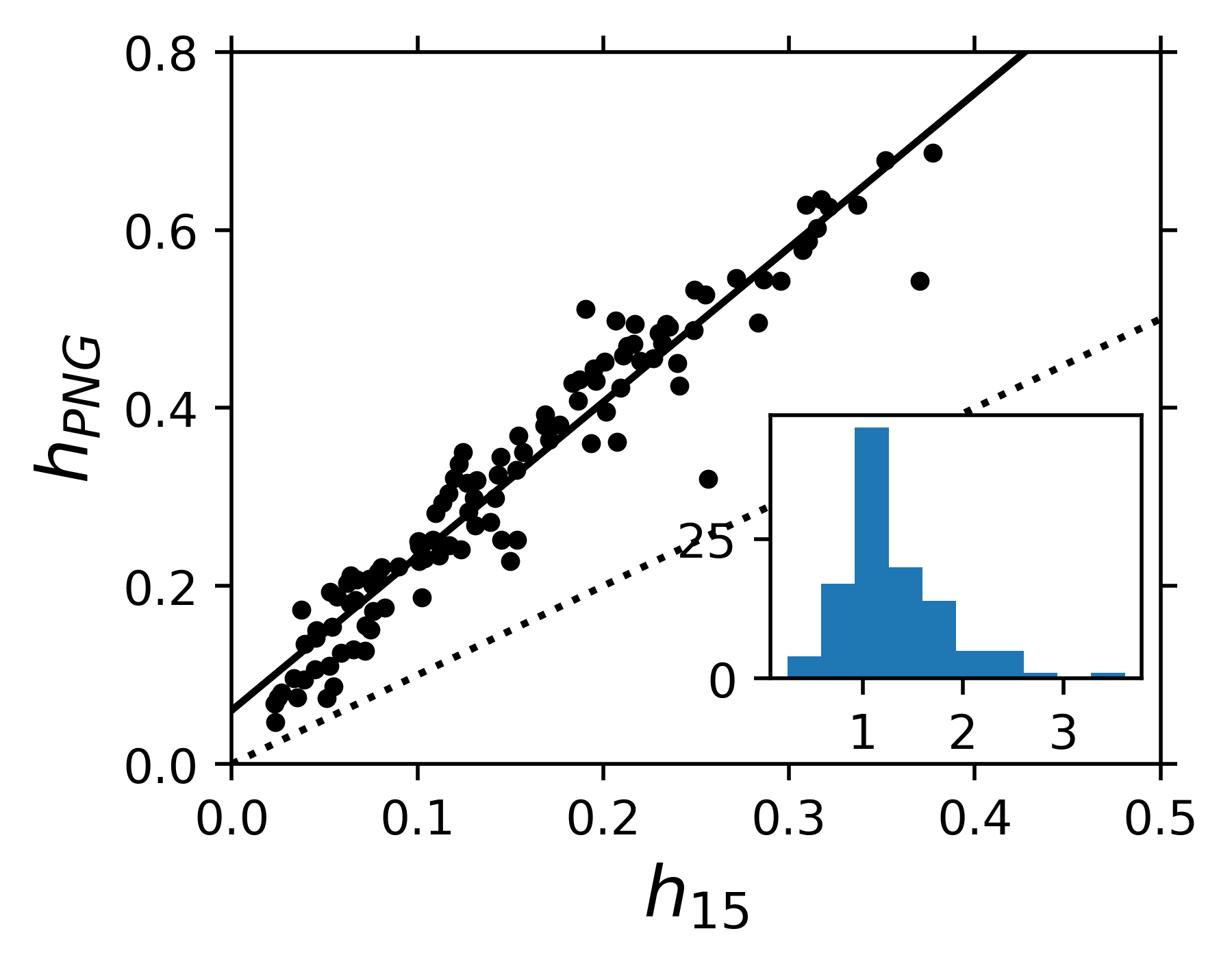}
\includegraphics[width=0.2\textwidth, angle=0]{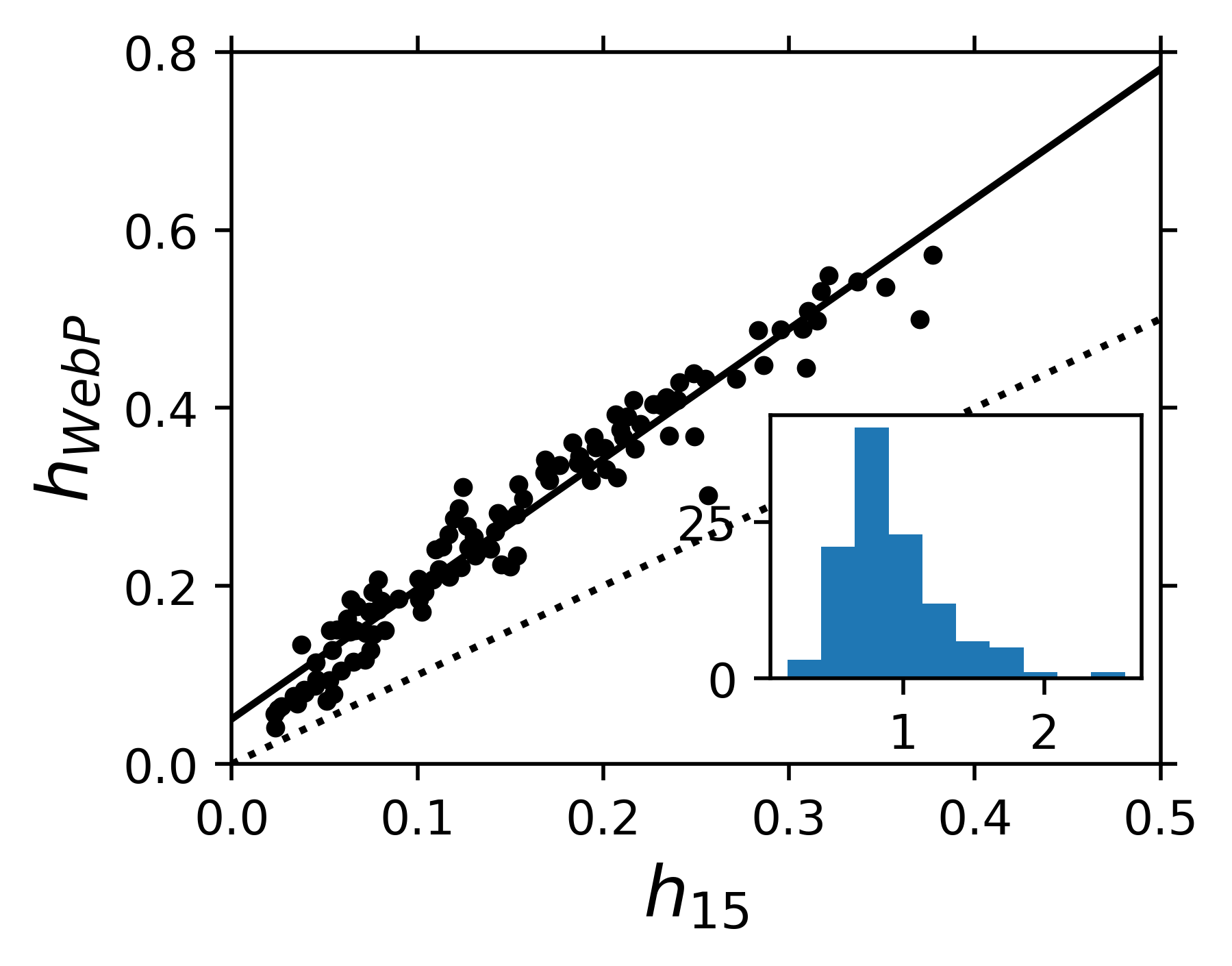}
\includegraphics[width=0.2\textwidth, angle=0]{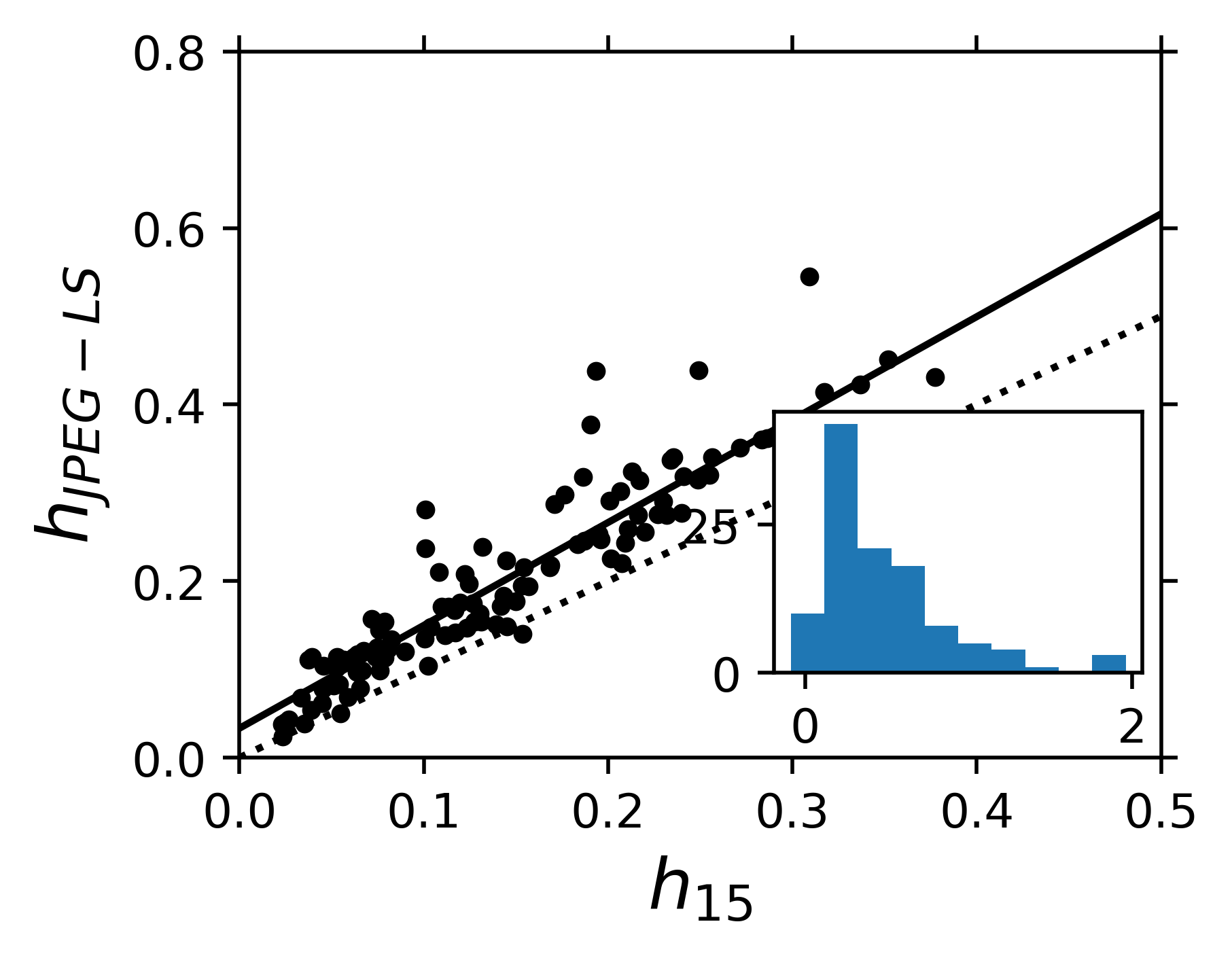}
\end{center}
\caption{\small Scattering plot of the entropy of the Brodatz textures evaluated 
using the compression algorithms versus the block-entropies algorithm. 
Continuous lines represent  the linear fitting of the points. In the inset,   
the histogram of the relative differences. 
}
\label{fig_Brodatz}
\end{figure}

We systematically compare the entropy estimated by using the 
block-entropies method of equation \ref{hentropy2} 
with the ones measured with the compression method
for different lossless compression algorithms ($h_{c.a.}$).
For the block-entropies method we consider the results obtained for
blocks up to order 15 ($h_{15}$).
This value assures that we avoid relevant finite size effects,
as proven by the fact that for a perfectly random binary matrix the method,
at this order, underestimates the entropy at about 2\%.


Comparison between the two methods is depicted in
Figure \ref{fig_city}  for the data of urban sections, and in
Figure \ref{fig_Brodatz} for Brodatz's set of textures.
Figures represent the scattering plot of the pairs ($h_{15},h_{c.a.}$), where $c.a.$
is the name of the applied compressor algorithm.
It is evident that, in general, the  compression method dramatically overestimates 
the entropy. The only exception is the Jpeg-ls algorithm, which presents
a closer estimation of the entropy to the values of $h_{15}$.
These results are characterized using a linear fit of the scattering plot
and evaluating the percentage difference: 
\begin{equation}
r=\frac{h_{c.a.}-h_{15}}{h_{15}}.
\label{eq_diff}
\end{equation}
In general, linear fittings show that the difference between 
$h_{c.a.}$ and $h_{15}$ grows for larger $h_{15}$ values. 
Considering the two datasets, the median of the percentage difference $r$ is between 75\% and 135\%
for all the considered compressors, excepted the Jpeg-ls which presents a median of 36\% 
and a very weak dependence of the difference between $h_{c.a.}$ and $h_{15}$ on the entropy value. 
Details about this analysis can be found in Table \ref{table1}, both  for the urban sections data and Brodatz's textures.

\begin{table}
    \centering
{\small
    \begin{tabular}{|c|c|c|c|c|}  
\hline 
Compressor & $\bar{r}$ & $\tilde{r}$ & a & b \\
\hline 
\hline
		 Gzip  &
		 0.98  &
		 0.91  &
		 1.25  &
		 0.15 \\
		 Png  &
		 1.08  &
		 1.03  &
		 1.41 &
		 0.14 \\
		 WebP  &
		 0.80  &
		 0.75  &
		 1.37 &
		 0.09 \\
		 Gif  &
		 1.12  &
		 1 .05  &
		 1.09 &
		 0.21  \\
		 Jpeg 2000  &
		 1.12  &
		 1.07  &
		 1.26 &
		 0.18  \\
		 Jpeg-ls  &
		 0.42  &
		 0.36  &
		 0.88 &
		 0.11  \\
\hline
\hline





\hline 
\hline

		 Gzip   &
		 1.17  &
		 1.05  & 
		 1.63  &
		 0.06 \\
		 Png  &
		 1.30  &
		 1.19  &
		 1.73  &
		 0.06 \\
		 WebP  &
		 0.93  &
		 0.83  &
		 1.46  &
		 0.05 \\
		 Gif  &
		 1.29  &
		 1.15  &
		 1.68  &
		 0.07 \\
		 Jpeg 2000  &
		 1.46  &
		 1.35  &
		 1.71  &
		 0.08 \\
		 Jpeg-ls  &
		 0.48  &
		 0.36  &
		 1.17  &
		 0.03 \\
		\hline
\end{tabular}
}
\caption{ Results of the analysis of the difference
in the entropy estimation between the block-entropies and the compression method.  
$\bar{r}$ is the mean value of the measured $r$ and $\tilde{r}$ the median.
The first table contains the results obtained from the urban sections;  
$a$ is the slope and $b$ the y-intercept  of the linear regressions of Fig. \ref{fig_city}. 
 The second one shows the results obtained from Brodatz's textures; 
 $a$ and $b$  results from the linear regressions of Fig. \ref{fig_Brodatz}.
}
\label{table1}
\end{table} 

\begin{figure}[h]
\begin{center}
\vspace{0.6cm}
\includegraphics[width=0.4\textwidth, angle=0]{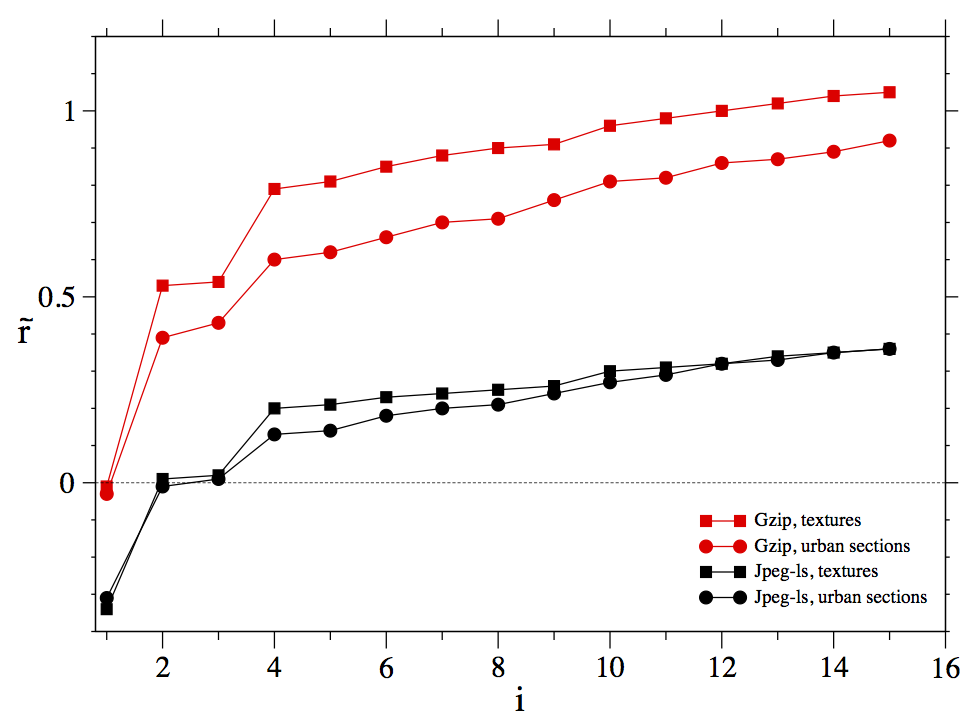}
\end{center}
\caption{\small 
$\tilde{r}$ values displayed in dependence of the order of the block-entropies
method used in the comparison with the compression method implemented by using the Gzip (red points) and Jpeg-ls algorithms (black points) for the urban sections (circles) and Brodatz's textures (squares).}
\label{fig_size}
\end{figure}

In the following, we focus our analysis to the two most representative methods of 
compression: the LZ77 algorithm  implemented by Gzip with Hilbert's curve
and the Jpeg-ls algorithm.
The first because it has been used and explored most frequently, 
the second because it displays, by far, the best performance among the considered compressors.
In Figure \ref{fig_size} we plot the median of $r$ ($\tilde{r}$) as a function of the order $i$
of the considered block-entropies method for these two compression methods.
The Gzip performance is comparable to the order-1 block entropies method 
and the Jpeg-ls algorithm approximates
the results of the order-3 block entropies.
Up to order 4 there is an abrupt improvement in the performance of the block-entropies method. 
Then, a linear growth follows. 

\begin{figure}[h]
\begin{center}
\vspace{0.6cm}
\includegraphics[width=0.35\textwidth, angle=0]{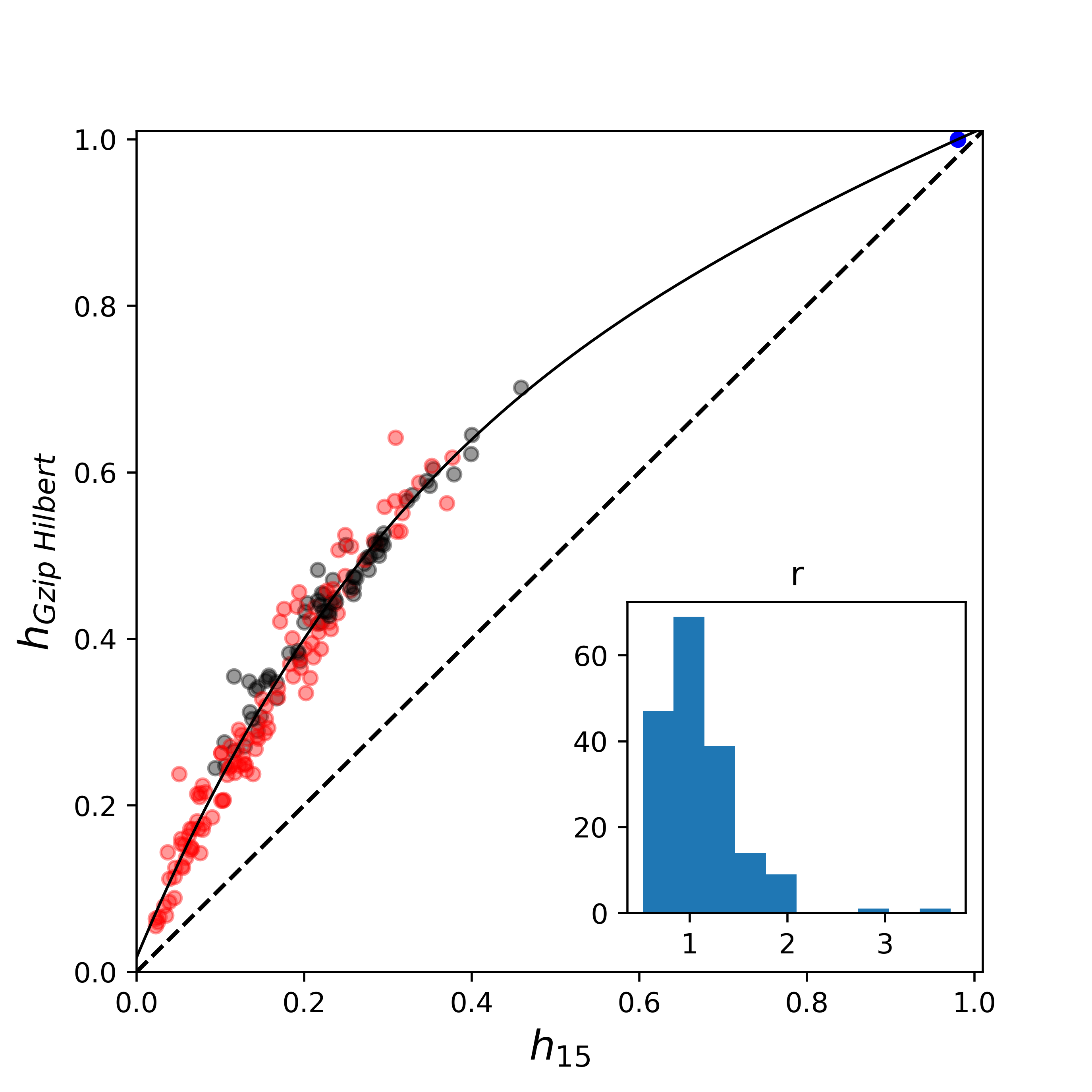}
\includegraphics[width=0.35\textwidth, angle=0]{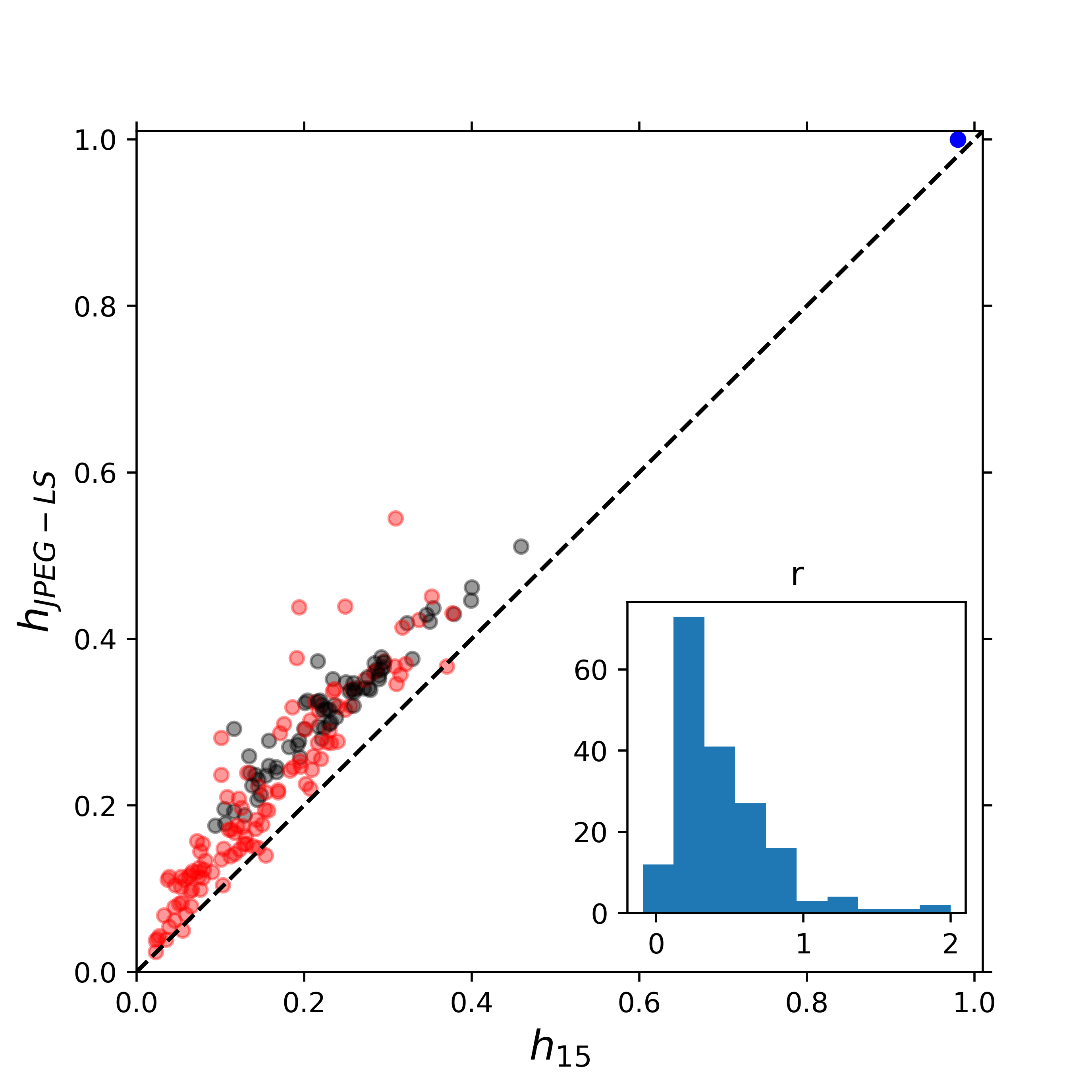}
\end{center}
\caption{\small Scattering plot of the entropies evaluated 
using the compression algorithm versus the block-entropies algorithm. 
Red points represent data of Brodatz's textures, black points of 
the urban sections. The blu point is the result of a perfectly random image.
In the inset,  the histogram of the relative differences 
between the two considered methods.  The dashed line is the $x=y$ equation.
 {\it Top}: the Gzip compression algorithm is used.
The continuous line is a guide for the eyes (generated by a 
4-th order polynomial fitting).
{\it Bottom}: results for the Jpeg-ls compression algorithm.
}
\label{fig_all}
\end{figure}

Figure \ref{fig_all} represents the comparison of the methods 
using the full pool of data, consisting of the urban sections and Brodatz's textures.
It outlines how the different approaches perform over all the considered spectrum of entropies.
For construction, all methods give exactly 0 for a perfectly uniform image.
The compression framework gives 1 for a perfectly
random set (for construction), and the block-entropies approach gives practically 1
for all $h_i$ with $i\le15$.
For this reason, the scattering plot shows a natural convergence 
towards these values and, consequently, an equivalence between the two methods
at the extremes of the spectrum.
In contrast, for intermediate values the comparison shows
a clear increase in the overestimation of the entropy for the compression methods.
This fact is particularly evident for the Gzip algorithm implementation.
These entropy values correspond to all the data from urban sections, which practically 
always present long-range correlations.
In this range of entropy  values the two datasets practically fully overlap.

By looking at the results of the Jpeg-ls compressor we can note
that the dependence of the comparison on $h$ is weaker
and, in general, points are closer to the $x=y$ line.
On the other hand, the pairs show a pronounced spreading.
Focusing on the points which present
an important overestimation of the entropy measured with
the compression method,
we note that they correspond to textures characterized by some
symmetries along a specific axe.

\begin{figure*}[h]
\begin{center}
\vspace{0.6cm}
\includegraphics[width=0.45\textwidth, angle=0]{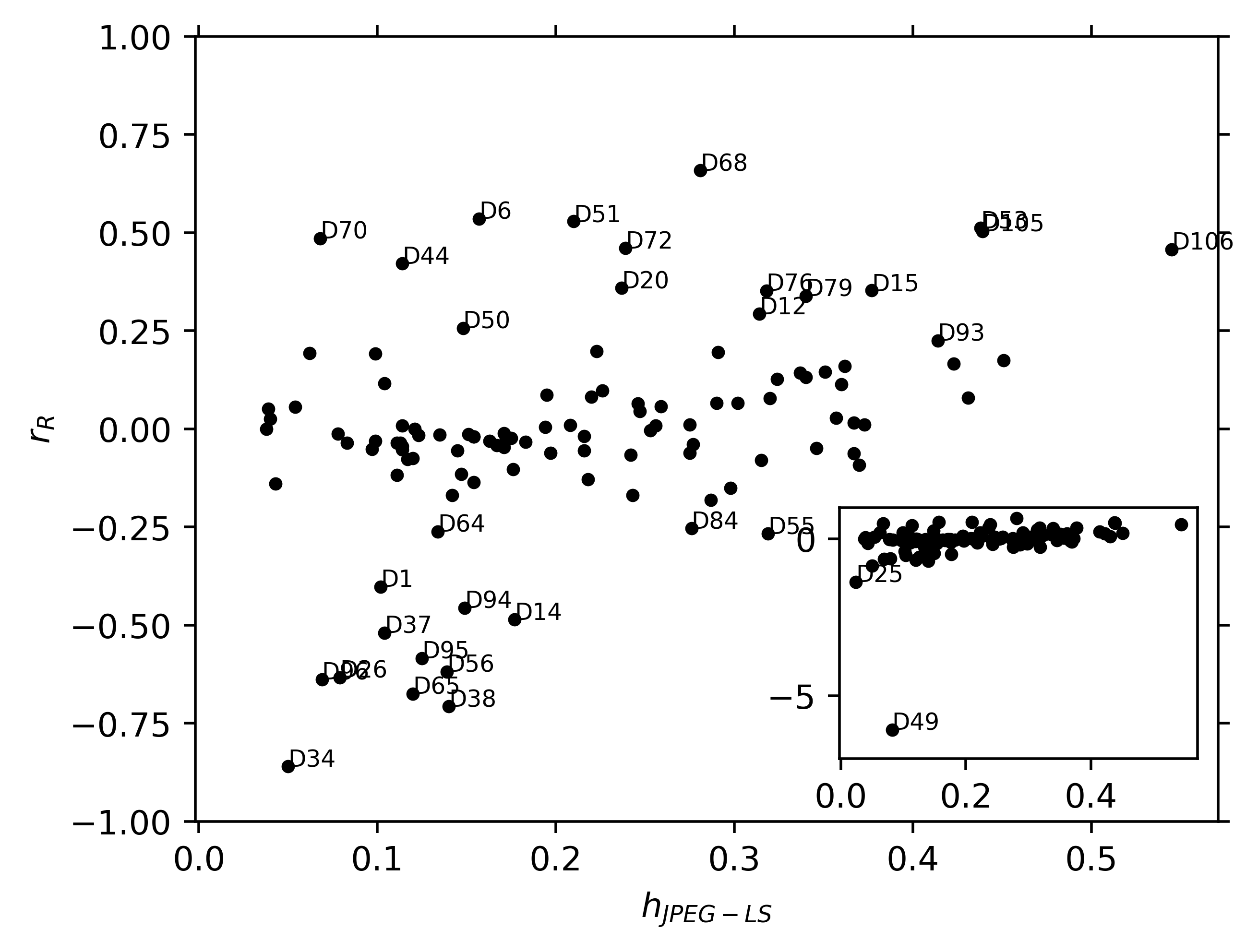}
\includegraphics[width=0.45\textwidth, angle=0]{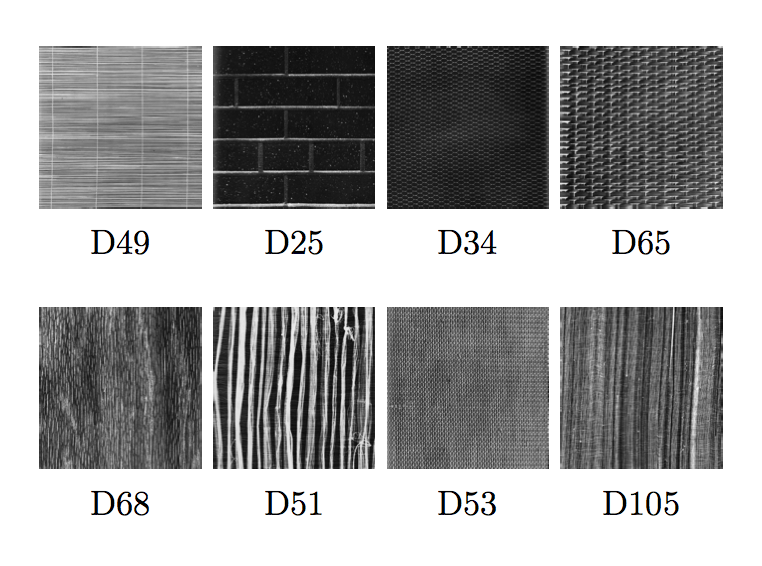}
\includegraphics[width=0.45\textwidth, angle=0]{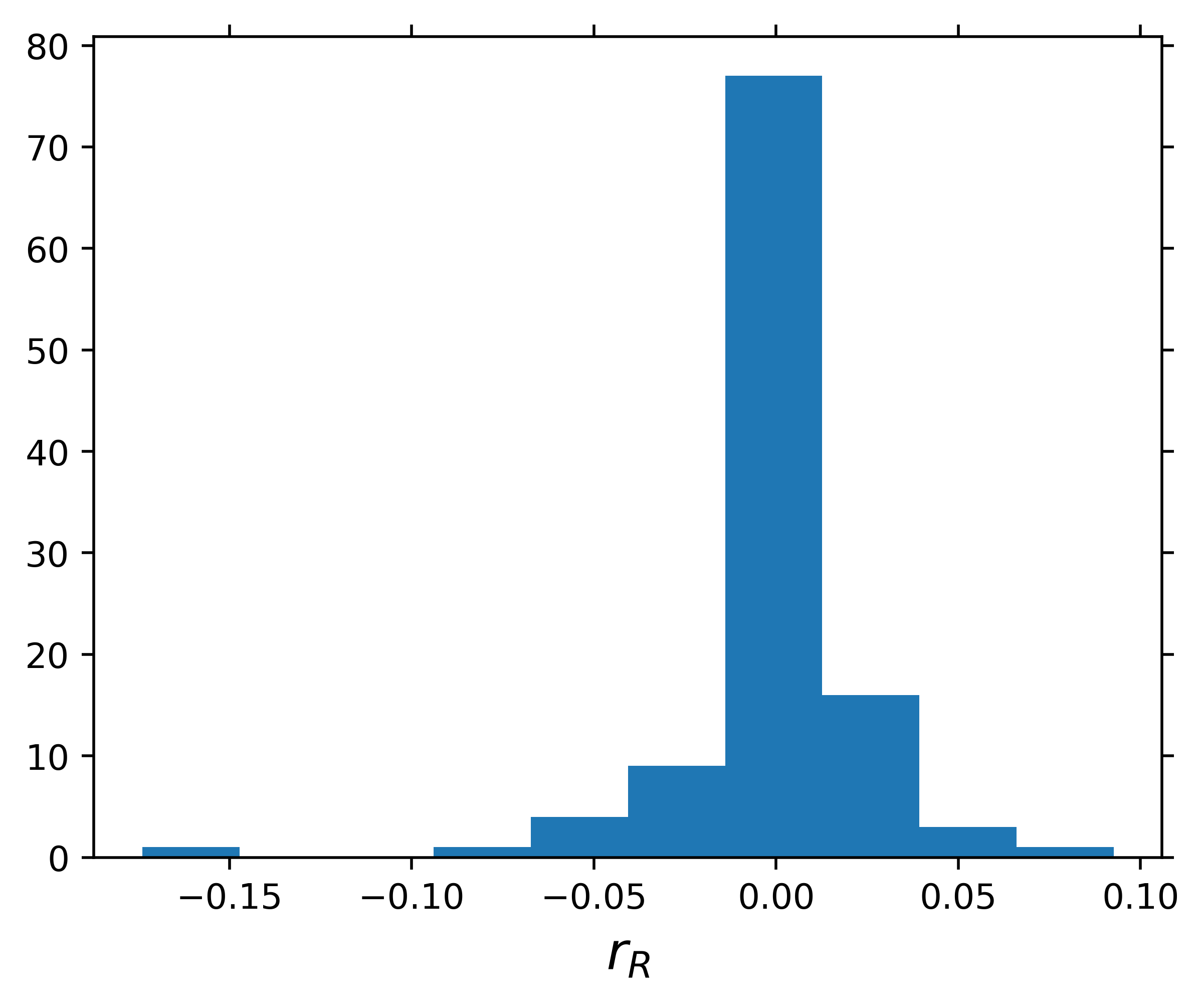}
\includegraphics[width=0.45\textwidth, angle=0]{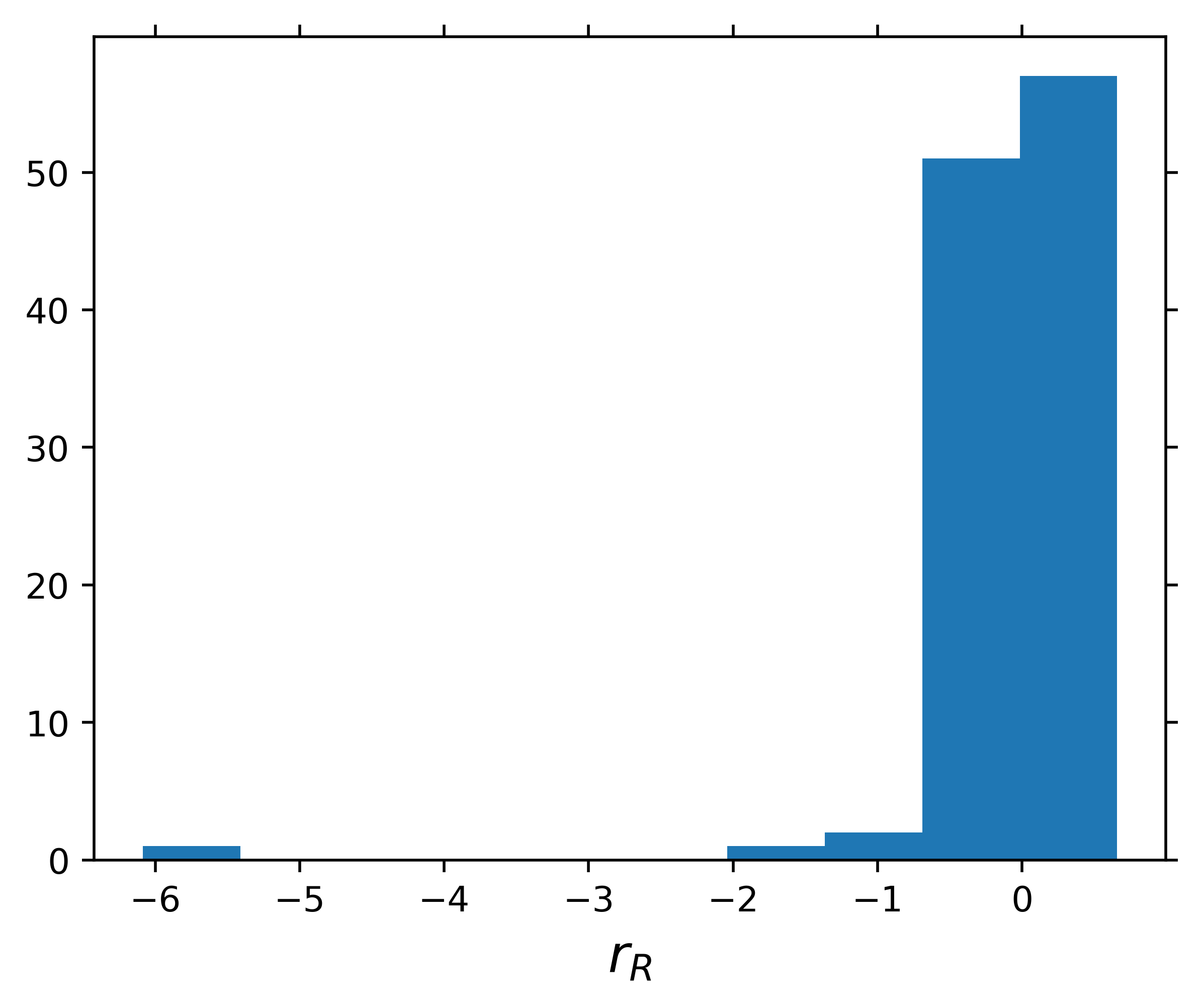}
\end{center}
\caption{\small 
{\it Top}: on the left, the relative difference between the entropy estimated from the original and rotated  
Brodatz's textures using the Jpeg-ls compressor method. 
The textures with $|r_R|>0.2$ are indicated with their symbols.
In the inset, the same data encompassing also the extreme statistics.
On the right,  
eight textures presenting large differences.
{\it Bottom}: distribution of the $r_R$ values for the Jpeg-ls compressor (on the left) and the 
block entropy method (on the right). The first distribution presents a standard deviation of 0.646,
the second one of only 0.028.
}
\label{fig_rota}
\end{figure*}

We explore this behavior produced by the Jpeg-ls method 
by estimating the entropy for the original Brodatz textures ($h^O$) 
and for the same ones after a $90\,^{\circ}$ rotation ($h^R$).
The relative difference of these measures, 
defined as $r_R=\frac{h^O-h^R}{h^O}$ is shown in Figure \ref{fig_rota}.
Note that all the images which present a high $|r_R|$ value display
line segments, stripes or structures predominantly oriented in horizontal 
or vertical directions.
That is all the most expected, since Jpeg-ls performs line-by-line sweeps using predictor blocks which are 3-pixels wide and 2-pixels tall, hence the existence of vertical or horizontal patterns will benefit the compression of one of the rotated images in a more pronounced fashion, whereas images without those axis-aligned patterns tend to be more uniformly processed by Jpeg-ls, irrespective of rotation. 
In contrast, if the block-entropies method is used, the $r_R$ is systematically very small
and can be ascribed to noise effects.
This independence of the block entropy method from rotations has been
previously verified in \cite{Brigatti21}, using random scanning paths.

\section{Discussion}

The comparison of the block-entropies method with the framework based on lossless
compressors shows that this second approach, independently of the considered compressor
algorithm, always overestimates substantially the entropy.  
The algorithms Gzip, Png, WebP and Gif present similar results,
with the median of the percentage differences enclosed in the interval $[0.75,1.19]$.
This means that the overestimation of the entropy for these approaches, 
in relation to the block-entropies method of order 15, is around 100\%.
The analysis of the dependence of this difference on the order of the block-entropies method
shows that blocks with a size of only 4 elements are sufficient for obtaining 
evident better estimations than the compression algorithms, which seem able to grasp correlations 
only up to the first neighbors cells (see Figure \ref{fig_size}).
The algorithms Gzip, Png, WebP and Gif are based on the idea of projecting the 2D sequences  
onto a lower dimensionality and then processing these strings using the LZ scheme, generally 
implemented by the Deflate algorithm.  
It seems that the specific details introduced by each compressors,
either in preprocessing the images or in coding,
often optimized for colored or gray sequences,
do not have a relevant impact on the final results, 
at least for binary sequences.

We conjecture that the principal obstacle for reaching better estimations is 
the 1D projection of the sequences. 
The reduction of 2D patterns onto a 1D string significantly destroys involved bidimensional structures, 
which are particularly significant in systems with long-range correlations, and generates a substantial 
overestimation of the entropy.
A hint in this direction comes from the use of a conceptually different compressor, the Jpeg-ls,
which is based on inference and uses bidimensional paths. This algorithm
strongly reduces the overestimation producing a percentage difference of 36\%.
This fact shows that the framework introduced in \cite{Avinery} is dependent
on the type of the implemented compression scheme, 
where not only the compression algorithm but also the eventual use of a projection 
procedure is relevant.  
Moreover, it shows that the use of algorithms specifically designed for bidimensional systems 
can substantially improve the performance of the approach. 
If, on one hand, the underestimation is still important and the results surely can not be 
considered accurate for general 2D systems, on the other, this approach
can be interesting for specific situations. 
For example, when we are interested in rough and fast estimations or in 
relative entropy values and when the sequences are not binary. 
In this last situation, the block-entropies 
method becomes ineffective, as for estimating the probabilities of countless block configurations
a huge statistic should be necessary.

Our analysis produced other interesting results.
By considering a large pool of 180 images with very different characteristics, we can describe how the different approaches perform over a larger spectrum of entropies. Figure \ref{fig_all} shows the expected correct convergence towards 0 and 1 at the extremes of the spectrum. In contrast, for intermediate values, the comparison clearly shows the overestimation of the entropy for the compression method. 
These results well represent the 
dependence of the quality of the estimates 
on the specific entropy values \cite{Lesne}, 
outlining another well-known aspect which makes entropy estimation difficult.
Finally, the dependence of the performance of the compression method implemented with 
Jpeg-ls on the presence of axial symmetries makes this method a potentially interesting algorithm 
for automatic symmetries detection in textures and images. 
In fact, 
if a particular rotation exists which produces an evident lower value in the detected entropy, the horizontal direction of this rotated image corresponds to the axis where patches (stripes, segment or other patterns) are predominantly oriented. 

To sum up, we tested the convergence properties of 
two methods of entropy estimation 
for 2D binary sequences with a fixed typical size. 
We showed that frameworks based on lossless
compression methods applied to data projected 
onto a one-dimensional string lead to poor estimates. 
This is because higher-dimensional 
correlations are obscured by the projection operation.
The adoption of compression methods which do not realize 
the dimensionality reduction ({\it i.e.} Jpeg-ls) can improve the 
performance of this approach.
Traditional block-entropies methods
generalized to 2D systems show a really faster convergence 
and clearly better results in estimating the asymptotic entropy.
 



\section{Conclusion}

The literature studying entropy estimation in bidimensional systems
is very sparse, and some aspects of the applied methods 
have generated controversy.
Recent works suggest that lossless compression algorithms 
can be used as an efficient and accurate approach 
to solve this task universally,
proposing that 2D sequences may be systematically 
linearized by appropriate procedures \cite{Avinery,Martiniani}
and leading to the belief that such an approach can perform at least as well as the 
block-entropies method does \cite{Martiniani}.

Recently, it has been shown that traditional block-entropies methods 
are flexible and robust enough to be applied on general 2D 
systems \cite{Brigatti21}.
Here, our analysis demonstrated that the use of this approach 
allows for more precise estimation of the asymptotic entropy 
than do frameworks based on lossless compression methods,
which cannot be considered accurate for 2D long-ranged correlated systems.
Moreover, in contrast to suggestions of previous works,  
an important dependence on the algorithms used by the compression methods
is outlined. Although general-purpose and algorithms designed explicitly
for compressing 
images can produce similar results, the Jpeg-ls specific code outperforms these
algorithms.
Some of these results may not seem surprising, yet they are presented here, in part to fill a gap in the literature and, in part, because they may lead to further insight into how entropy estimation of an individual 
image can be ultimately performed.\\

The potential applications of these results are very wide.
Entropy and free-energy calculations are fundamental for the thermodynamic analysis of general 2D  many-body systems at equilibrium. 
Furthermore, these techniques can be useful for the characterization of order and correlations in out-of-equilibrium systems and in other areas of statistical physics and dynamical systems.

Shannon's entropy is the 
cardinal quantity for developing 
information-based measures, with a direct interpretation in terms of 
disorder and a rigorous one in term of surprise.
It is a global unparameterized quantity, more sensitive and general
than traditional two-point measures, such as standard correlations. 
For these reasons, it can be used for 
characterization, diagnosis, modeling and classification of images and general spatial systems
({\it e.g.} medical and biological images, 
surfaces, turbulence, geographical systems and landscape patterns).
For example, 
the estimated entropy 
of the adopted dataset of urban sections has been already 
used for classifying these systems 
in \cite{Brigatti21,Netto18}.
Similar ideas could be applied for image characterization of 
Brodatz's textures.

\section*{Acknowledgments} 

F.N.M.S.F. received partial financial support from the Pibic/UFRJ-CNPq program.
E.B. received financial support from the National Council for Scientific 
and Technological Development - CNPq (
Grant 305008/2021-8).
The authors acknowledge Vinicius M. Netto and Caio Cacholas for making data on 
urban sections available for our analysis.



\section*{Appendix A: Block entropies method}

\begin{figure}[h]
\begin{center}
\vspace{0.6cm}
\includegraphics[width=0.2\textwidth, angle=0]{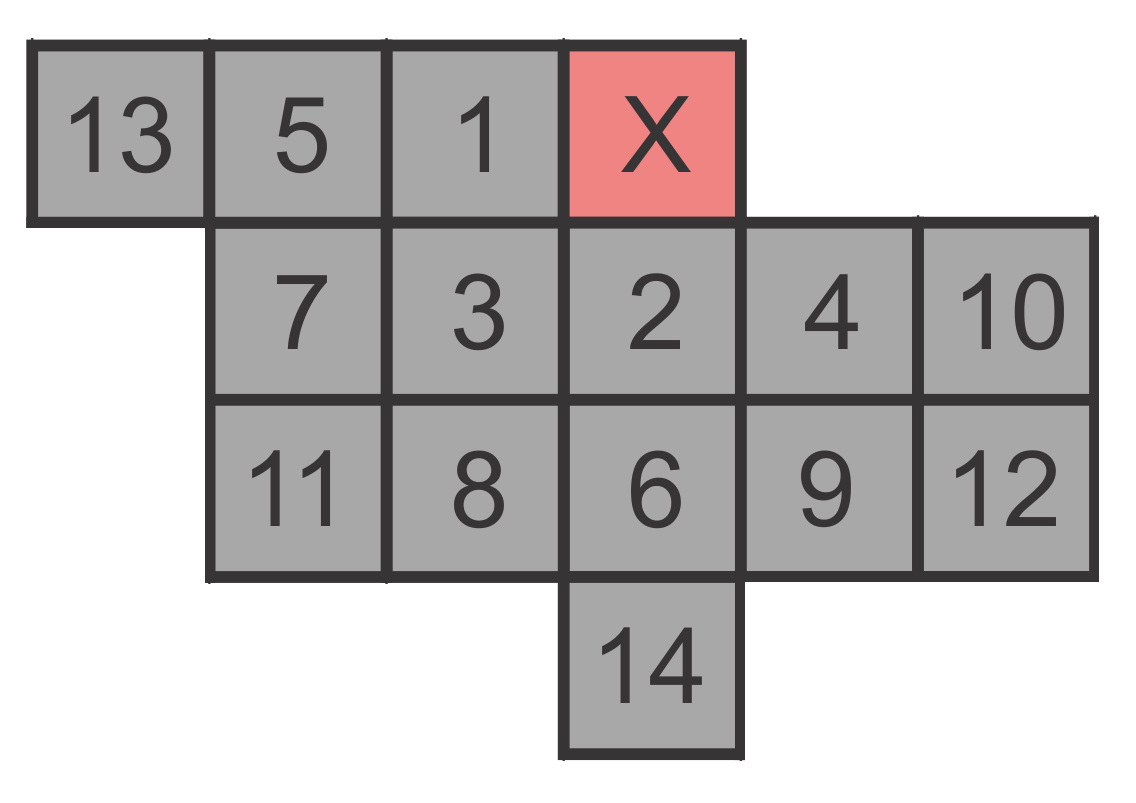}
\end{center}
\caption{\small This figure represents the block of size 15 for estimating $H_{15}$. $X$ is the target cell and
numbers indicate the scan order for general blocks up to order 15. 
These blocks have been clearly defined and described by Feldman {\it et al.} in \cite{Feldman03}.
}
\label{fig_block}
\end{figure}

We estimated Shannon's entropy as the limit of the differential entropies $h_n = H_n - H_{n-1}$.
This difference measures the randomness generated by adding the target cell (denoted by $X$ in Fig.\ref{fig_block}) to the block, given that we have already observed the cells in $H_{n-1}$.
We approximated this limit with the maximum order which can avoid relevant finite size effects; 
for this reason we calculate $h_n$ up to order 15 ($h_{15}$).
To do this, we must estimate $H_{15}$ and $H_{14}$. 

$H_{15}$ is obtained using the equation:

\begin{equation}
H_{15}=-\sum_k  p_{15}(k) \log_{2}[p_{15}(k)].
\label{1entropyBIS}
\end{equation}

We partitioned our matrix of $N\times N$ cells with $S$ blocks of size $15$, following the block geometry shown in Fig.~\ref{fig_block}.
All the $k$ possible different block configurations are taken into account and 
the number of times each configuration $k$ is found in the matrix ($Y_k$) is counted. 
The probability $p_{15}(k)$ is approximated by $Y_k/S$. 
By using these probabilities the 15-block entropy value $H_{15}$ is calculated using equation \ref{1entropyBIS}.
The value of  $H_{14}$ is obtained analogously. 
In this case, the block of Fig.~\ref{fig_block} is considered without the target cell $X$.

\section*{Appendix B: Lossless compression  method}

This method measures the compressed file size after the application of a lossless compression algorithm ($C_d$) and estimates the incompressibility $\eta$, which corresponds to the asymptotic entropy $h$.

$\eta=(C_d-C_0)/(C_1-C_0)$, where $C_0$ is a compressed degenerate datasets and $C_1$ 
a compressed random dataset.\\

\subsection*{General purpose compressor (Gzip):}

The 2D binary sequence is stored in a 1D file of size $N^2$. 
This is done scanning the 2D matrix with a Hilbert's curve.
The 1D file \verb+flat_matrix+ is compressed using:\\

\verb+ Z = gzip.compress(flat_matrix,9)+\\

and then the size of the zipped file $Z$ is measured, giving $C_d$.  
$C_1$ is obtained measuring the size of a compressed vector of size $N^2$ containing 
0's and 1's chosen independently and uniformly at random.
$C_0$ is obtained measuring the size of a compressed vector containing only 0.

\subsection*{Compressors designed for images (Png, WebP, Gif, Jpeg2000, Jpeg-ls):}

The 2D binary matrix is transformed into a bitmap B, substituting  all the 
 1 with 255 (white pixel) and leaving all elements 0 as 0 (black pixels). 
For example, in the case of the Jpeg-ls, the bitmap B is compressed using:

\begin{verbatim}
 with io.BytesIO() as image_file:
        image = Image.fromarray(B)
        image.save(image_file, 'jpeg-ls')     
\end{verbatim}

and then the size of the 
file \verb+image_file+ gives $C_d$.  
$C_1$ corresponds to the size of a compressed bitmap containing 
0 and 255, randomly sorted with equal probability.
$C_0$ is obtained measuring the size of a compressed 
bitmap containing only 0.


\end{document}